\documentclass[12pt,final,3p,times]{elsarticle}

\journal{Journal of \LaTeX\ Templates}

\usepackage{supertabular,booktabs}
\usepackage{longtable}
\usepackage{subfigure}
\usepackage{seqsplit}
\usepackage{hyperref}
\usepackage{caption}
\usepackage{amsmath}
\usepackage{blindtext}

\hypersetup{
    colorlinks=true,
    linkcolor=cyan,
    filecolor=cyan,      
    urlcolor=cyan,
}

\begin{document}

\begin{frontmatter}

\title{GoldEnvSim – A FLEXPART-WRF based software for simulation of radionuclides transport in atmospheric
\tnoteref{mytitlenote}}


\author[mymainaddress]{Nguyen Hong Ha}

\author[mysecondaryaddress,mymainaddress]{Phan Viet Cuong \corref{mycorrespondingauthor}}
\cortext[mycorrespondingauthor]{Corresponding author}
\ead{pvcuong@vinatom.gov.vn}

\author[mysecondaryaddress]{Le Tuan Anh}
\author[thirdadress]{Ho Thi Thao}
\author[fourthadress]{Hoang Huu Duc}
\author[fiveadress,sixthadress]{Kieu Ngoc Dung}

\address[mymainaddress]{Institute of Physics, Vietnam Academy of Science and Technology, Hanoi, Vietnam}
\address[mysecondaryaddress]{Research and Development Center for Radiation Technology, Vietnam Atomic Energy Institute, Vietnam}
\address[thirdadress]{School of Mechanical Engineering, Kyungpook National University, Korea}
\address[fourthadress]{Centre for Environment Treatment Technology, High Command of Chemistry, Hanoi, Vietnam}
\address[fiveadress]{National Committee for Search and Rescue, No 6, Phuc Dong, Long Bien, Hanoi, Vietnam}
\address[sixthadress]{Vietnam Atomic Energy Institute, 59 Ly Thuong Kiet Str, Hoan Kiem, Hanoi, Vietnam.}

\begin{abstract}
This article illustrates the development of a software named GoldEnvSim for simulation of the dispersion of radionuclides in the atmosphere. The software is written in JavaFX programming language to couple the Weather Research and Forecasting (WRF) model and the FLEXPART-WRF model. The highlight function of this software is to provide convenience for users to run a simulation workflow with a user-friendly interface. Many toolkits for post-processing and visualizing output are also incorporated to make this software more comprehensive. At this first version, GoldEnvSim is specifically designed to analyze and predict the dispersion of radioactive materials in the atmosphere, but it has potential for further development and applicable to other fields of environmental science. For demonstration, a simulation of the dispersion of the $^{137}$Cs that is assumed to be released from the Fangchenggang nuclear power plant to whole Vietnam territory was performed. The simulation result on meteorological in comparison to the monitoring data taken from a first-class meteorological observatory was used to evaluate the accuracy of dispersion simulation result.  
\end{abstract}

\begin{keyword}
GoldEnvSim, WRF, FLEXPART-WRF, Lagrangian dispersion model, Air dispersion modelling.
\end{keyword}

\end{frontmatter}

\onecolumn
\section*{Software availability}

• Name of software: GoldEnvSim.

• Developers: Phan Viet Cuong, Nguyen Hong Ha, Le Tuan Anh, Ho Thi Thao, Hoang Huu Duc.

• Address: Danang Branch, Research and Development Center for Radiation Technology, Hoang Van Thai street, Hoa Son commune, Hoa Vang district, Da Nang city, Vietnam.

• Contact: pvcuong@vinatom.gov.vn 

• First version available: 2019.

• Minimum hardware requirements: CPU Intel Core i3-6100, 8GB of RAM or equivalent.

• OS requirements: Tested on Ubuntu 16.04 x64 LTS, Ubuntu 18.04 x64 LTS. 

• Program language: Java FX 11.

• Program size: 20 MB for the software, 80 GB on disk for accompanying dependencies after installation.

• Cost: Undetermined at this time.

\section{Introduction}

Air dispersion simulation is a powerful technique to evaluate behaviors of air pollutants disperse in the ambient atmosphere. It is based on computer programs that contain models and algorithms to solve mathematical equations that describe dispersion processes and atmospheric phenomena. Numerical models are used under specific scenarios to simulate the consequences of air pollutants, hazardous chemical materials, or radionuclides released from industrial facilities or nuclear power plants. Some typical softwares/systems for air quality assessments as the ModOdor of Yanjun et al. \cite{1} which is an Eulerian-based model for simulating the dispersion of gaseous contaminants; the Openair of Carslaw et al. \cite{2} for air quality assessment. For fast nuclear accident response could be mention as the ENSEMBLE of Bianconi et al. \cite{3} for the management of long-range atmospheric dispersion of radioactive materials; the READY web-based of Glenn et al. \cite{4} for running the HYSPLIT model. These softwares have always still been successful in the fields where they born to serve, however the atmospheric dispersion modeling is a complex problem, therefore it needs to make more improvements in the dispersion models itself and accompanied softwares.

The matured of dispersion models passed through endeavors of development, the first-generation models based on Gaussian diffusion model accounted for the transportation on steady-state approximations in time and space, described the correlation of the horizontal and vertical dispersion to the observed standard deviations of lateral and elevation angles of the wind fluctuations \cite{5} \cite{6}. The second-generation models added removal processes, increased the level of sophistication in the parameterizations and chemistry simulations, allowed transport and dispersion as a function of time and space \cite{5}. The third-generation models, also known as the new-generation models, have been implemented with an advanced approach to describe diffusion and dispersion using the fundamental properties of the atmosphere rather than relying on general mathematical approximation \cite{7}, adopt more sophisticated features to provide turbulence parameters for estimating diffusion rates by using boundary layer and surface energy flux parameterizations \cite{8}. Moreover, they coupled together with new generation meteorological models make it successfully applied over complex terrain and regional scales. 

There are three ubiquitous types of mathematical models for simulating atmospheric dispersion: Gaussian, Eulerian, and Lagrangian. In the Gaussian model, the pollutant concentration profile through the plume follows a Gaussian distribution in vertical and lateral directions. The Gaussian model is mostly recommended for applications that need fast calculation, such as online risk management \cite{9}. The Eulerian model tracks the movement of a large number of pollution plume parcels as they move away from their initial location by solving the Navier-Stoke equation, it uses a fixed three-dimensional Cartesian grid as a frame of reference \cite{8}. The Eulerian model is well suited for long-range dispersion applications, or high stability of boundary layers \cite{9}. The Lagrangian model followed moving air parcels (also called particles) along trajectories, it calculates the velocity of the particles as a stochastic process \cite{7}. It is the most proposed model for applications that need to scrutinize on accuracy and simulation time in mesoscale, such as free atmospheric dispersion and complex terrain \cite{9}.

The FLEXPART is a Lagrangian dispersion model, it is an open-source program ever since it was released and has been written in Fortran 90 code. It is designed for calculating long-range and mesoscale dispersion of air pollutants released from point sources, such that occurring after an accident in a nuclear power plant. FLEXPART has been comprehensively validated with experiments by using controlled tracer in intercontinental air pollution transport studies \cite{10} \cite{11} \cite{12} \cite{13} \cite{14} \cite{15}. FLEXPART uses input data from global meteorological models provided by the European Centre for Medium-Range Weather Forecasts (ECMWF) or the National Centers of Environmental Prediction (NCEP-GFS) \cite{16}. However, the coarse resolution of the GFS/ECMWF data is low, this affects the planetary boundary layer (PBL) turbulence parameterizations which is valid for small-scales and local-scales \cite{17}. Besides, the coarse resolution of these data uses unrealistically smooth terrain which leads to an underestimation of orographic influences on precipitation \cite{18}. These factors will affect the accuracy in calculating the dispersion of radioactive in small-scales, which require high resolution of meteorological data.

The FLEXPART-WRF was developed to overcome these drawbacks of FLEXPART in small-scale applications, it uses input from the Weather Research and Forecasting (WRF) model. In this case, the WRF plays as an intermediary processor between the GFS data and FLEXPART-WRF. Currently, the newest version of FLEXPART-WRF software is 3.3.2 which combines the main characteristics of the FLEXPART and counting new features, including new options for using different wind data (e.g. instantaneous or time-averaged winds), output grid projections, computing efficiency \cite{17}. The FLEXPART-WRF has been extensively used for air quality studies or predict the dispersion of trace gases and aerosols from point, line, or area sources \cite{19}. A novel scheme for skewed turbulence in the convective PBL has been implemented based on the formulation developed by Cassiani et al., which may give significant improvements, especially for small-scale and complex terrain applications \cite{17}. It can illustrate the process of mesoscale transport and diffusion, computes dry/wet deposition of the tracer by forward simulation or determines potential source area through backward simulation. The transport equation consists of a resolved wind and a random turbulent component parameterized through the Langevin stochastic differential equation wherein the Lagrangian time scale, which is determined from boundary layer parameters (surface fluxes, mixing height, Monin–Obukhov length, convective velocity scale, roughness length and friction velocity) derived from the WRF meteorological model \cite{20}. 

The WRF is a numerical weather prediction model designed for both research and operational applications. WRF is supported as a common tool for the university/research and operational communities to promote closer ties between them and to address the needs of both \cite{21}. It is optimized for various forecast and analysis applications, from the microscale to the synoptic and even global scales. WRF includes dozens of parameterizations for boundary layer processes, convection, microphysics, radiation, land surface processes, and several options for numerical schemes \cite{17}. 

In the present, there is no GUI software coupling the FLEXPART-WRF and WRF to make a comprehensive software for atmospheric dispersion assessment. The idea to coupling of meteorological models with pollutant dispersion models was proposed by Mandurino et al. \cite{22}; Oza et al. with the ADOCT software \cite{23}; and Lasse et al. \cite{24}. These studies showed good reliability in terms of combining the meteorological models and dispersion models to predict pollutant dispersion in the atmosphere. For meteorological applications with WRF, there are several GUI software bespoke for WRF simulation, such as the WRF Domain Wizard developed by NOAA's Earth System Research Laboratory \cite{25}, and the QGIS by Meyer et.al \cite{26}. They satisfy users by conveniently defining simulation domains, selecting WRF schemes, run WPS (geogrid, ungrib, and metgrid) and WRF through the GUI, and visualize the NetCDF output. The FLEXPART-WRF is powerful as it brings, but there is no powerful GUI software to provide an easy way to run it, users have still been configuring it by setting up every single variable in a text file, and run it by order input manually. Therefore, the objective of this work is to develop a GUI software named GoldEnvSim to couple the WRF and the FLEXPART-WRF together, using for simulation the dispersion of radionuclides in small-scales and mesoscales. It provides convenience for users when configuring parameters for simulation, it will take place to serve for both research purpose or real-time forecast the dispersion of radioactive material in specific scenarios. The highlight functions of this software are operating on the user interface and automatically linking the output of WRF to FLEXPART-WRF. In this article, we introduce our software GoldEnvSim and some results on the calculation of the dispersion of $^{137}$Cs released from the Fangchenggang nuclear power plant to Vietnam territory for demonstration. This manuscript is organized into four sections. \hyperref[sec:2]{Section 2} illustrates the details of the software, its components and functions. \hyperref[sec:3]{Section 3} provides simulation results on the dispersion of the $^{137}$Cs assumed to be released from the Fangchenggang nuclear power plant to whole Vietnam territory. Lastly, \hyperref[sec:4]{Section 4} gives main conclusions on this research and further perspective for development of this software.

\section{Materials and methods} \label{sec:2}
\subsection{General overview} \label{sec:2.1}
The developed software GoldEnvSim aims to provide a GUI platform to support users to run FLEXPART-WRF simulations in an easy way. GoldEnvSim can be running in three main functions, which serve for each particular purpose: Fast response accident (Quick setup function); Real-time simulation function; and Advanced setup function. These functions will be discussed further in \hyperref[sec:2.3]{Section 2.3}. In addition, several visualizers such as Google maps, IDV, Quicklook are also integrated into this software to display simulation results.

The GoldEnvSim has been written in JavaFX 11, the minimum system required for operating is CPU Intel Core i3-6100, 8GB of RAM, 80 GB of storage on disk after installation and operating platform is Ubuntu 18.04 x64 LTS or higher. 

\subsection{Installation} \label{sec:2.2}
Required dependencies for running WRF and FLEXPART-WRF could be installed automatically by a package we provide. By executing a bash script file under superuser privilege (\textit{sudo}), required dependencies will be installed. We have enclosed following packages in the installer:

\begin{table}[]
\caption{Required dependencies composed in the GoldEnvSim installer.}
\begin{tabular}{|l|l|}
• Oracle Java SE 11. & • HDF5 1.8.12.        \\
• Gfortran 4.9.1.    & • IDV 5.4.            \\
• Gcc 4.4.0.         & • Numpy 1.17.0.       \\
• Cpp 4.8.4.         & • Matplotlib 2.2.2.   \\
• Python 2.7.        & • Basemap 1.2.1.      \\
• NetCDF 4.1.3.      & • Basemap Data Hires. \\
• Mpich 3.0.4        & • Quicklook 1.0.      \\
• Zlib 1.2.7.        & • WPS 4.0.            \\
• Libpng 1.2.50.     & • WRF 4.0.            \\
• Jasper 1.900.1.    & • FLEXPART-WRF 3.3.2.
\label{tab:tab1}
\end{tabular}
\end{table}

\subsection{Description of components and software development} \label{sec:2.3}

\begin{figure*}
\centering
  \includegraphics[width=0.9\linewidth, height=2.8in]{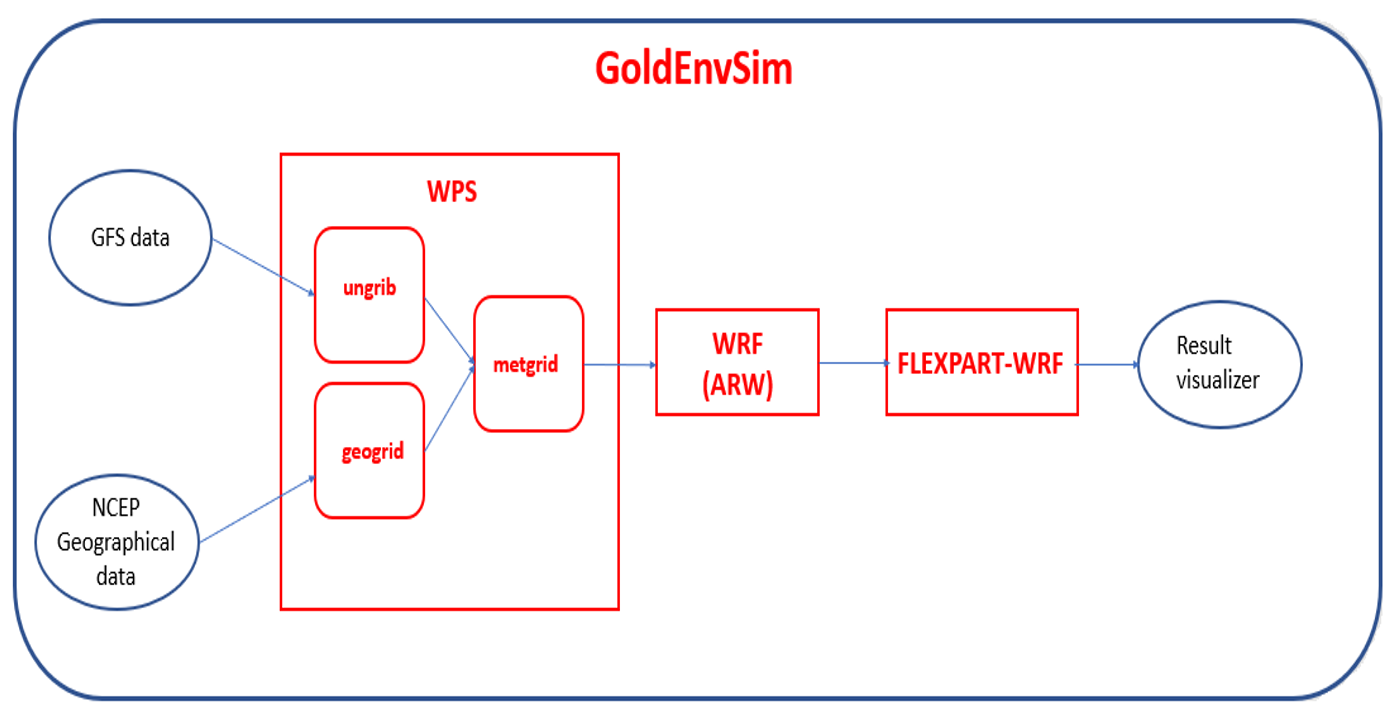}
  \caption{Schematic of the GoldEnvSim components.}
  \label{fig:fig1}
\end{figure*}

GoldEnvSim consists of three main modules: WPS (The WRF Pre-processing System), WRF, and FLEXPART-WRF. \autoref{fig:fig1} shows the schematic of its structure.

WPS contains three sub-modules: \textit{geogrid}, \textit{ungrib}, and \textit{metgrid}. The \textit{geogrid} program mapping computational domain, defines geographical projection, sets domain resolutions, and creates static files of terrestrial data. The \textit{ungrib} program decodes the NCEP GFS files in GRIB format for meteorological data. And the \textit{metgrid} program horizontally interpolates the meteorological data onto the projected domains \cite{21}. The output data from WPS in NetCDF format supplies a complete three-dimensional snapshot of the atmosphere on the selected model grid’s horizontal staggering at the selected time slices \cite{21}.

This software embedded the Google Maps Javascript API in the interface to define meteorological domains and emission position. The Google map is accurately for determining the longitude/latitude of objects in the map, therefore it is bespoke to determine the coordinate of the center point and corner points of simulation domains. The distance between two corner points of each domain is calculated based on their coordinates by the Haversine formula \cite{27}, these distance results are used to calculate grid point number of the simulation domain. All information about longitude/latitude of calculation domain achieved from Google map will be stored in a variable and wrote to the WPS configuration file (\textit{namelist.wps}). GoldEnvSim executes a bash script file contains commands to run \textit{geogrid}, \textit{ungrib}, \textit{metgrid} of WPS, and link netCDF output files of WPS to WRF.

The WRF consist of 2 dynamic solver cores: NMM (Non-hydrostatic Mesoscale Model) and ARW (Advanced Research WRF). Recently, the funding for development of NMM has no longer been supported \cite{28}. The ARW is a non-hydrostatic mesoscale model based on Eulerian (mass conservative) dynamical core \cite{20}. In this software, we used the ARW core for meteorological purpose and simply invoke it as WRF. 

The WRF contains two programs, the \textit{real} program takes input from WPS to vertically interpolate and establish lateral boundary conditions \cite{21}. And the \textit{wrf} program for numerical integration the forecast model by using physics parameterization schemes, these schemes are configured in the user-defined file \textit{namelist.input}. For config the WRF simulation, GoldEnvSim allows users to import the \textit{namelist.input} file which already contains preset about WRF schemes, or users can select WRF schemes on the interface of the software and save this setup as a user-defined configuration file for later uses. GoldEnvSim runs the \textit{real} and \textit{wrf} programs by executes a bash script file that contains the command: 
\begin{small}
\begin{verbatim}
mpirun -np [number of cores] [name of program]
\end{verbatim}
\end{small}

When the WRF module finishes running, its output will be automatically loaded to FLEXPART-WRF module to calculate the dispersion of radionuclides or other kinds of particles. 

As a flexible model, FLEXPART-WRF is configured by the \textit{flex.namelist.input} file which is contains setups about wind conversion, boundary layer treatment, release location and the type of species assume to be released. The crucial role of the GoldEnvSim is sync between output of WRF as input configuration of FLEXPART-WRF, we have implemented a sub-routine in the GoldEnvSim by using \textit{ncdump} command to automatic extract information of WRF netCDF output files to get latitude/longitude value of domain corners and the time step of simulation, then store in variables to write to the \textit{flex.namelist.input} file. The wind option is fixed to be snapshot wind; the turbulent option for handle PBL uses the turbulent kinetic energy (TKE) from WRF; the calculations for fluxes and convection are all switched on.

In the FLEXPART-WRF module, users need to define the emission area, the release start/end time, emission rate, and the content of source terms. All of these manipulations can be performed via the interface of the software. At this first version, the list of species in the GoldEnvSim includes $^{85}$Kr, $^{90}$Sr, $^{91}$Y, $^{106}$Ru, $^{131}$I, $^{133}$Xe, $^{137}$Cs, all of them are in aerosol form. GoldEnvSim contains information on decay half-life of particles, molar weight, the Henrys constant for describing the solubility of the gas, the average diameter of particles, the in-cloud scavenging and below-cloud scavenging coefficients of particles. This information is illustrated on the \autoref{tab:tab2}.
 
\begin{table*}[]
\centering
\caption{Information of the species defined in the GoldEnvSim software. \cite{16}}
\label{tab:tab2}
\begin{tabular}{|l|l|c|l|l|l|l|l|l|l|l|l|}
\hline
\multicolumn{1}{|c|}{Species} & \multicolumn{1}{c|}{\begin{tabular}[c]{@{}c@{}}Half \\ life \\ (sec)\end{tabular}} & \begin{tabular}[c]{@{}c@{}}Mole \\ weight \\ (g/mol)\end{tabular} & \multicolumn{1}{c|}{\begin{tabular}[c]{@{}c@{}}A\\ ($s^{-1}$)\end{tabular}} & \multicolumn{1}{c|}{B} & \multicolumn{1}{c|}{D} & \multicolumn{1}{c|}{\begin{tabular}[c]{@{}c@{}}Henry \\ (M/atm)\end{tabular}} & \multicolumn{1}{c|}{f0} & \multicolumn{1}{c|}{\begin{tabular}[c]{@{}c@{}}rho\\  (kg.$m^{-3}$)\end{tabular}} & \multicolumn{1}{c|}{\begin{tabular}[c]{@{}c@{}}dquer \\ (mm)\end{tabular}} & \multicolumn{1}{c|}{\begin{tabular}[c]{@{}c@{}}dsig \\ (mm)\end{tabular}} & \multicolumn{1}{c|}{\begin{tabular}[c]{@{}c@{}}vd \\ (m/s)\end{tabular}} \\ \hline
$^{85}$Kr                         & 3.39E08                                                                            & 85                                                                & 1.0E-04                                                                & 0.8                    & -9.9                   &                                                                               &                         & 2.5E03                                                                       & 6.0E-07                                                                    & 3.0E-01                                                                   & -9.9                                                                     \\ \hline
$^{90}$Sr                         & 9.08E08                                                                            & 90                                                                & 1.0E-04                                                                & 0.8                    & -9.9                   &                                                                               &                         & 2.5E03                                                                       & 6.0E-07                                                                    & 3.0E-01                                                                   & -9.9                                                                     \\ \hline
$^{91}$Y                          & 5.06E06                                                                            & 91                                                                & 1.0E-04                                                                & 0.8                    & -9.9                   &                                                                               &                         & 2.5E03                                                                       & 6.0E-07                                                                    & 3.0E-01                                                                   & -9.9                                                                     \\ \hline
$^{106}$Ru                        & 3.23E07                                                                            & 106                                                               & 1.0E-04                                                                & 0.8                    & -9.9                   &                                                                               &                         & 2.5E03                                                                       & 6.0E-07                                                                    & 3.0E-01                                                                   & -9.9                                                                     \\ \hline
$^{131}$I                         & 6.93E05                                                                            & 131                                                               & 1.0E-04                                                                & 0.8                    & -9.9                   &                                                                               &                         & 2.5E03                                                                       & 6.0E-07                                                                    & 3.0E-01                                                                   & -9.9                                                                     \\ \hline
$^{133}$Xe                        & 4.53E05                                                                            & 133                                                               & 1.0E-04                                                                & 0.8                    & -9.9                   &                                                                               &                         & 2.5E03                                                                       & 6.0E-07                                                                    & 3.0E-01                                                                   & -9.9                                                                     \\ \hline
$^{137}$Cs                        & 9.49E08                                                                            & 137                                                               & 1.0E-04                                                                & 0.8                    & -9.9                   &                                                                               &                         & 2.5E03                                                                       & 6.0E-07                                                                    & 3.0E-01                                                                   & -9.9                                                                     \\ \hline
\end{tabular}
\end{table*}

Where:

• A ($s^{-1}$) :  scavenging coefficient for wet deposition at precipitation rate is 1 mm/hour. Off if A$<$0. \cite{16}

• B: dependency on precipitation rate. Both A and B are used to calculate scavenging coefficient $\Psi = A$×$I^B$, where I is precipitation rate. \cite{16}

• D: molecular diffusivity of gases in air, for calculate the quasi-laminar sublayer resistance of the dry deposition of gases by Erisman et al. theory. For gases D$>$0, for aerosol D$<$0. \cite{16} \cite{30}

• Henry (M/atm): Henry’s constant, describing the solubility of the gases in dry deposition calculation. \cite{30}

• f0: Chemical reactivity factor for oxidation of biological substances compare to ozone. For non-reactive species f0 is 0, for slightly reactive species it is 0.1 and for highly reactive species it is 1. \cite{31}

• rho (kg.$m^{-3}$): Particle density for dry deposition. For gases rho$<$0, for aerosol rho$>$0. \cite{16}
dquer (mm): Particle mean diameter $\overline{d}$, for gases $\overline{d}\leq$0, for aerosol $\overline{d}>$0 \cite{30}

• dsig (mm): Particle diameter standard deviation. \cite{30}

• vd (m/s): Dry deposition velocity of gases. For gases vd$>$0, for aerosol vd$<$0. \cite{16}

GoldEnvSim can be used on super-computers, cluster or PCs due to running FLEXPART-WRF in multi-threading mode by using MPI libraries. To run the FLEXPART-WRF, this software execute following command:
\begin{small}
\begin{verbatim}
./flexwrf33_gnu_mpi [path to the current project folder
/flex.namelist.input] 
\end{verbatim}
\end{small}

The output of WRF is in netCDF format as multiple raster layers which contains dozen meteorological fields, such as wind, pressure, temperature, specific humidity, rain…In order to visualize these outputs, we implemented the IDV (Integrated Data Viewer) for post-processing of meteorology simulation. It is a Java-based free software framework for analyzing and visualizing geoscience dataset, especially in netCDF format \cite{32}

To visualize the output of FLEXPART-WRF, we use the Quicklook toolkit which developed by Radek Hofman \cite{33}. The Quicklook is coupled with the Google maps and the Matplotlib-Basemap for display trajectory of species on map. Quicklook handles each FLEXPART-WRF output files, exports as a PNG image by timestep, and then merges them into a GIF animation image. Users can observe the concentration trajectory of radioactive plume released from emission area by animation images, the dry deposition and wet deposition is also demonstrated on. The Quicklook toolkit is automatically executed after the FLEXPART-WRF module finish. 

\subsection{Software features} \label{sec:2.4}
As mentioned above, GoldEnvSim has three main functions for running simulation: Quick setup simulation; Real-time simulation; and Advanced setup simulation. Each mode is suitable for each demand of users, for briefly, the “Quick setup simulation” is served for fast config of WRF schemes; the “Real-time simulation” allows download the latest GFS data from NCEP; the “Advanced setup” allows users to select WRF schemes and customs release scenarios. All details of these modes are described below.

\begin{figure*}
  \includegraphics[width=\linewidth, height=3.5in]{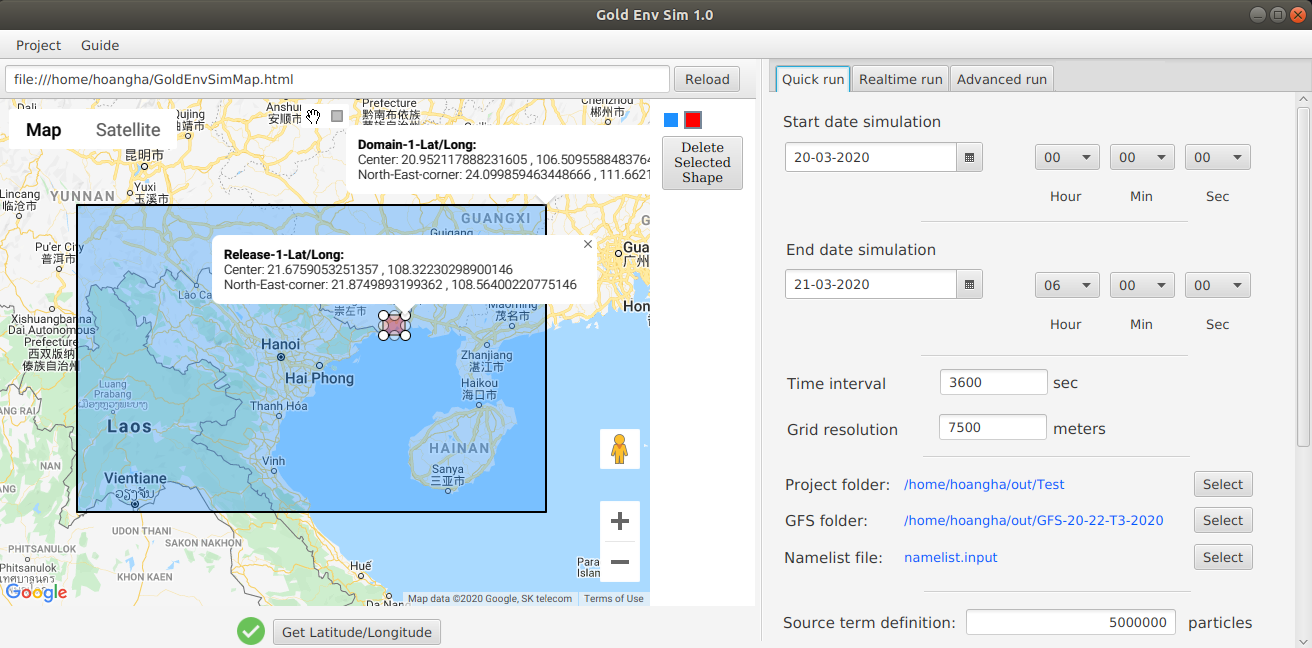}
  \caption{Interface of the GoldEnvSim software.}
  \label{fig:fig2}
\end{figure*}

\begin{longtable}{|p{0.45\linewidth}|p{0.45\linewidth}|}


\hline \hline 
\endfirsthead 

\hline \hline
\endlastfoot

\textbf{a} \includegraphics[width=0.98\linewidth, height=10.0cm]{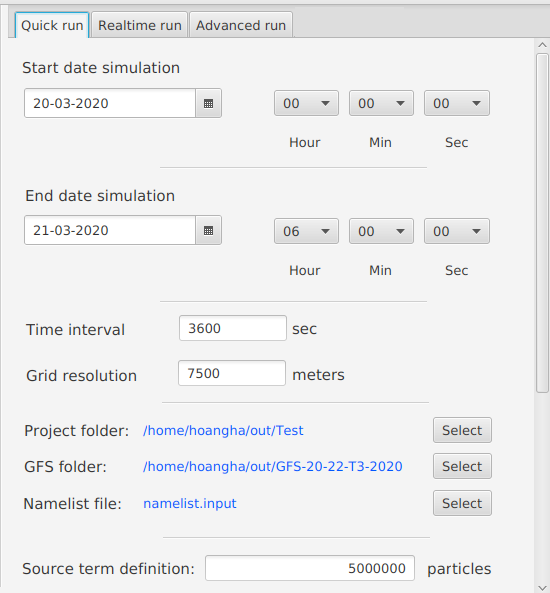} \label{fig:fig3a} & \textbf{b} \includegraphics[width=0.98\linewidth, height=10.0cm]{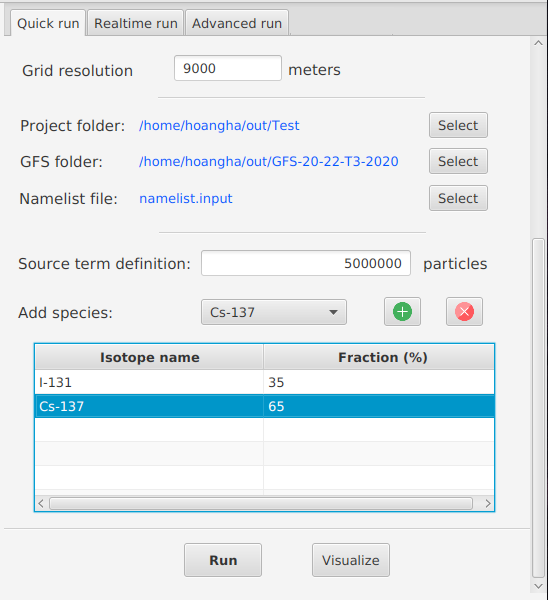} \label{fig:fig3b} \\ \hline 

\textbf{c} \includegraphics[width=0.98\linewidth, height=10cm]{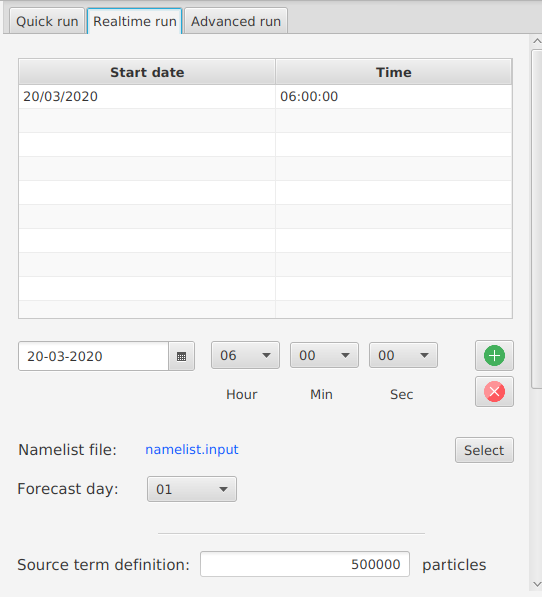} \label{fig:fig3c}
&\textbf{d} \includegraphics[width=0.98\linewidth, height=10cm]{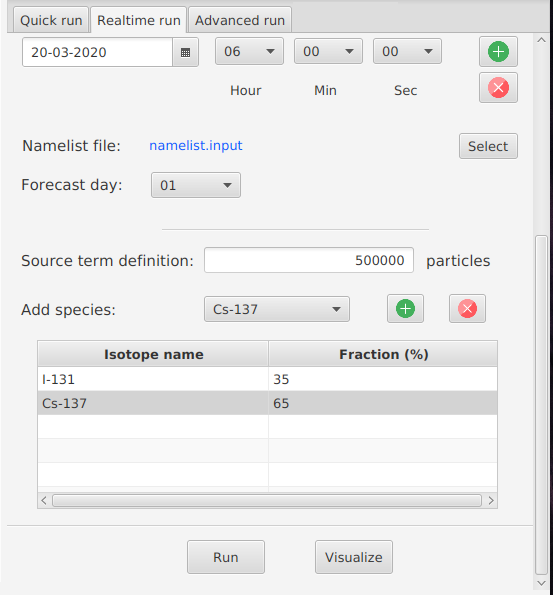} \label{fig:fig3d}  \\ \hline

\textbf{e}\includegraphics[width=0.98\linewidth, height=10cm]{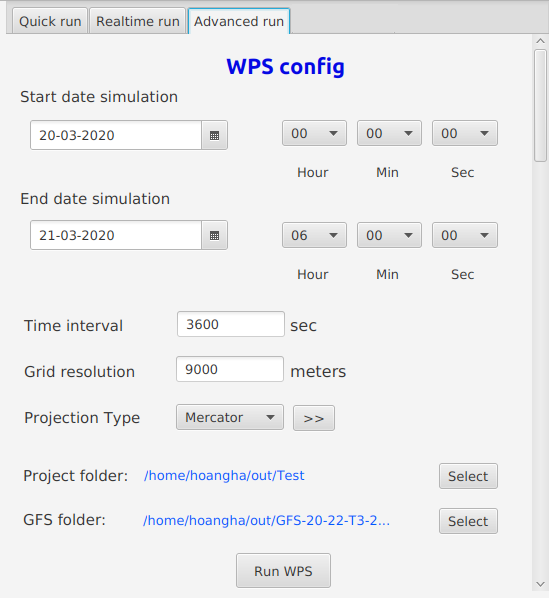} \label{fig:fig3e}
& \textbf{f} \includegraphics[width=0.98\linewidth, height=10cm]{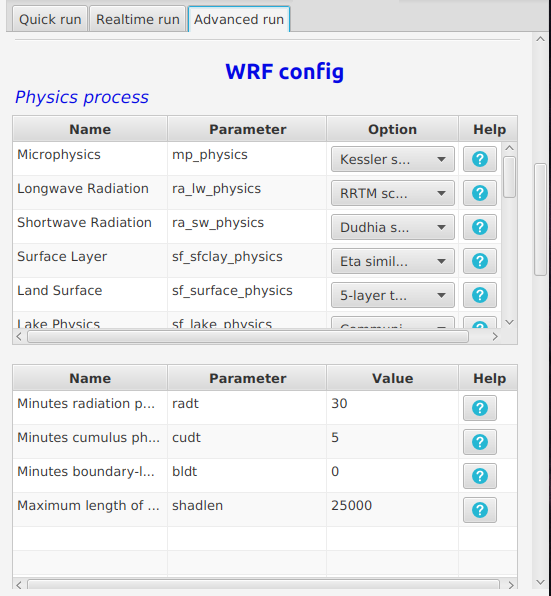} \label{fig:fig3f}   \\ \hline 

\textbf{g} \includegraphics[width=0.98\linewidth, height=10cm]{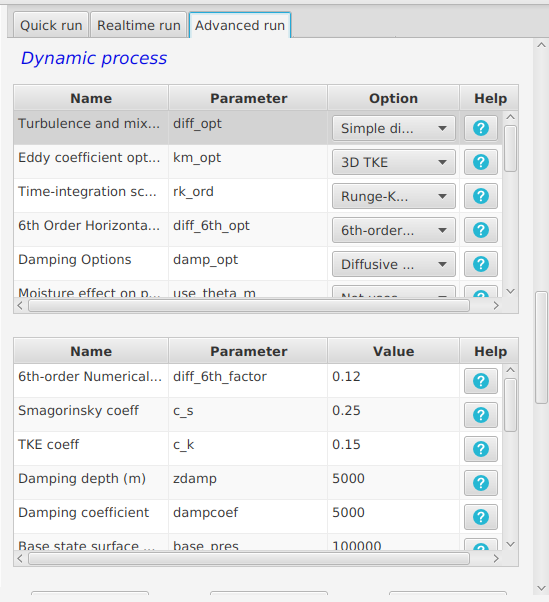} \label{fig:fig3g}
& \textbf{h} \includegraphics[width=0.98\linewidth, height=10cm]{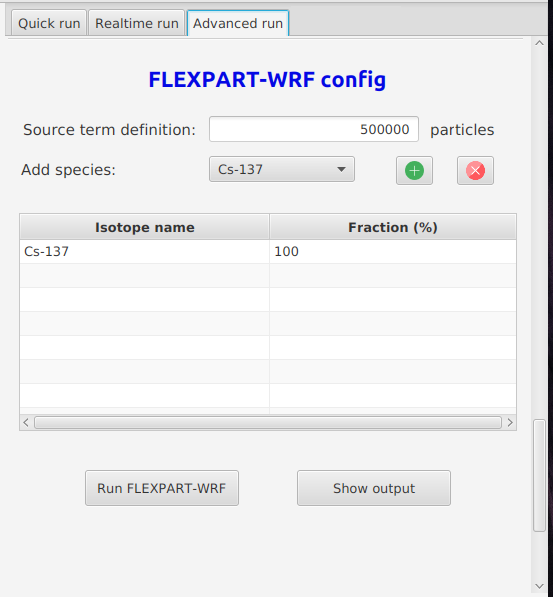} \label{fig:fig3h}  \\ \hline 
\end{longtable}
\captionof{figure}{Overview of the main features of the GoldEnvSim software.} \label{fig:fig3}

\subsubsection{Quick setup simulation} \label{2.4.1}
The “Quick setup” subtab (\autoref{fig:fig3}a and \autoref{fig:fig3}b) provides the easiest way to launch a simulation. The “Quick setup” means users can save time in config WRF schemes by loading a preset \textit{namelist} file which contains WRF schemes instead of selecting them one by one. Users also need to define the simulation period by setting up the start/end date of simulation, prepare GFS data for simulation, setting up time step interval, grid resolution, and source term components. The information required for setting up the source term components is the total number of particles to be released, the fraction of each species, and the release rate in each time period.

The “Time step interval” represents for the time resolution or time step of the WRF output, users can define on the interface of the software, the default value is 3600 seconds for WRF. The time step of the FLEXPART-WRF output is fixed to default every 10 minutes. Users can also define the grid resolution of the mother grid on the software interface. A spacing ratio between gridpoint distance of the mother domain and sub-domain is fixed at 3:1, which is typical uses for Arakawa-C grid.

\subsubsection{Real-time simulation} \label{2.4.2}
In real-time meteorological forecast simulation using WRF, using the latest GFS data gives better accuracy than old data, this also means that dispersion simulation results will be more accurate when taking the latest GFS data. The “Real-time run” subtab (\autoref{fig:fig3}c and \autoref{fig:fig3}d) has a function that meets users demands in real-time forecasts of the dispersion of radioactive, it allows users to download the latest GFS data automatically from the NCEP server. NCEP provides GFS data four times per day on their website, to download these data, GoldEnvSim executes the \textit{wget} command to download GFS files from \href{ftp://ftp.ncep.noaa.gov/pub/data/nccf/com/gfs/prod/}{ftp://ftp.ncep.noaa.gov/pub/data/nccf/com/gfs/prod/}. In this feature, users need to define the start/end date of simulation, setting up time step interval, grid resolution, and source term components.

\subsubsection{Advanced setup simulation} \label{2.4.3}
The “Advanced run” (\autoref{fig:fig3}e to \autoref{fig:fig3}h) is revolved around advanced research purposes, allows users to select every single WRF scheme instead of loading them from a preset file. At each scheme option, we provide a help information about that scheme based on the WRF User’s Guides, this helps users more convenient in choosing WRF scheme. In this first version of GoldEnvSim contains physics and dynamic schemes of the WRF version 3.9.1. This subtab is helpful to see how influence of WRF schemes to the dispersion simulation results.

In this feature, users can run each GoldEnvSim’s module separately, from WPS, WRF, and FLEXPART-WRF, it required to set the start/end date of simulation, grid resolution, time step interval, and define source term components. Besides, users can also select three different types of projection that most suitable for each geographical location on the earth: Polar, Lambert, Mercator.

\subsubsection{Visualization} \label{2.4.4}

As described briefly in the \hyperref[sec:2.1]{Section 2.1}, GoldEnvSim uses the IDV for displaying the output of WRF, uses the Google map and Quicklook to visualize the the output of FLEXPART-WRF.

\begin{figure*}
  \includegraphics[width=0.9\linewidth, height=3.7in]{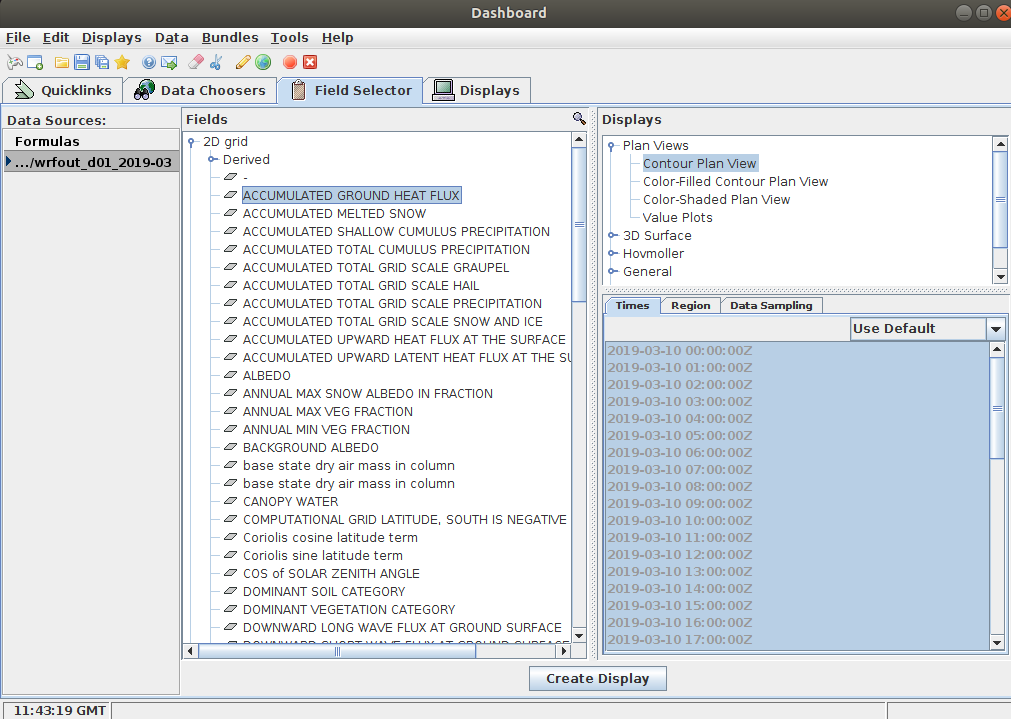}
  \caption{The IDV was integrated into the GoldEnvSim for WRF output dataset readout and visualization.}
  \label{fig:fig4}
\end{figure*}

In case of displaying WRF output, the IDV was integrated in the GoldEnvSim for visualization, user can visualize the meteorological output after GoldEnvSim done the WRF task. We developed a button to execute a command for opening IDV, users can access output sources of WRF after simulation and select fields to visualize. The IDV is able to render categorical data of the WRF output files, display them on map view, easily save them in many types of data such as GIF files for animation, MP4 video files, or KMZ files as a series of layers to display on the Google Earth. On the dashboard pane, users can select 2D or 3D field which exists in a WRF output file (\autoref{fig:fig4}). IDV provides a graphical environment for display datasets in diversity mode, including contour plan view; color-filled contour plan view; color-shaded plan view; or hybrid display of meteorological field and terrain field. For illustration, see \hyperref[sec:3]{Section 3} on a demonstration run using the GoldEnvSim and WRF output dataset visualization by the IDV.

To visualize the output of FLEXPART-WRF, the GoldEnvSim automatic executes commands to process binary output files of the FLEXPART-WRF after it has been done the simulation. This process is taken by the Quicklook which is developed by Radek Hofman \cite{33}. The Quicklook is written in pure python 2.7 and it needs Numpy and Mathplotlib-BaseMap as prerequisites. Data processed by the Quicklook are displayed in single PNG or JPEG files and then are merged into an animated GIF file. It can provide the trajectory of dry deposition, wet deposition, and concentration of every released species which are defined by users. Besides, the Google Maps JavaScript API is taken into account for visualization by displaying the animated GIF file as a raster layer. Users have full control on the Google Maps as well as tracking the trajectory of each species.

\section{Demonstration run} \label{sec:3}
\subsection{Simulation configuration} \label{sec:3.1}
We used the GoldEnvSim to simulate a released case of $^{137}$Cs from the Fangchenggang nuclear power plant (Fangchenggang NPP). Among NPPs operating in China, the Fangchenggang NPP is the nearest one to Vietnam, approximately 50 kilometers towards the east of the Vietnam border. The Fangchenggang NPP has two CPR-1000 units operating since 2016, other two units are under construction, and two units are planned \cite{34}. 

We provided a brief simulation to track the trajectory of radioactive material assumed released from the Fangchenggang NPP to Vietnam territory. The total activity of $^{137}$Cs assumed to be release was 10$^{6}$ Bq, the simulation took 50,000 particles to gives sufficient distribution. The latitude and longitude of the plant site is 21.694929 and 108.370895, respectively. The release started from 00:36 of 10/03/2019 to 06:00 of 11/03/2019. This simulation ran on the High-Performance Computing (HPC) with 88 threads of Intel Xeon E5-2699 ~ 2.2 GHz, it took 37 minutes for running and about 12 Gb of disk storage.

The WRF schemes for running this simulation and release property are described in \autoref{tab:table3}, these schemes are referenced from the study has been done by Raghavan et al. \cite{35} on regional climate simulations over Vietnam. Besides, we made a comparison between WRF simulation results with monitoring data at the Lang station (first-class meteorological observatory, latitude: 21.01, longitude: 105.48, locating in Hanoi – the capital of Vietnam) on the temperature field, wind field, surface pressure, and precipitation field. These basic factors directly affect to the dispersion process of air pollutants and could be used to evaluate the accuracy of the dispersion result.

\begin{longtable}{|p{0.15\linewidth}|p{0.25\linewidth}|p{0.35\linewidth}|}
\caption{WRF setup for meteorological simulation.} \label{tab:table3} \\

\hline \multicolumn{1}{|p{0.15\linewidth}|}{\textbf{Parameter}} & \multicolumn{1}{p{0.25\linewidth}|}{\textbf{Configuration scheme}} & \multicolumn{1}{p{0.35\linewidth}|}{\textbf{Properties}} \\ \hline 
\endfirsthead

\multicolumn{3}{c}%
{{\bfseries \tablename\ \thetable{} -- continued from previous page}} \\
\hline \multicolumn{1}{|c|}{\textbf{Parameter}} & \multicolumn{1}{c|}{\textbf{Configuration scheme}} & \multicolumn{1}{c|}{\textbf{Properties}} \\ \hline 
\endhead

\hline \multicolumn{3}{|r|}{{Continued on next page}} \\ \hline
\endfoot

\hline \hline
\endlastfoot

Initial boundary data file & GFS data forecast & Freely access. Widely uses for research and operation.  \\ \hline 

Meteorological & 214 grid point in West-East. 316 grid point in South-North. Centered at lat:16.17, long: 106.61. & Cover whole Vietnam terrority and Fangchenggang NPP.  \\ \hline

Grid resolution & 6 km &  \\ \hline 

Map projection & Mercator & Best used for tropical regions, especially with Vietnam.  \\ \hline 

Microphysics & Thompson scheme & Many improments in gamma distribution function for cloud water droplets parameterization; algorithms for calculation vapor deposition/sublimation and evaporation has been upgrade. \cite{21}\\ \hline

Cumulus Parameterization & Grell-Devenyi scheme & A mass-flux type scheme. The dynamic control closures provide ensembles of 144 members, these member are based on the convective available potential energy.  \cite{21}\\ \hline

Longwave \& shortwave radiation & Rapid Radiative Transfer Model scheme & A spectral-band scheme using the correlated-k method. \newline Uses pre-set tables to accurately represent longwave processes due to water vapor, ozone, CO2. \newline  Accounting for cloud optical depth. \cite{21}\\ \hline

Surface layer & MM5 similarity scheme & Based on Monin-Obukhov with Carslon-Boland viscous sub-layer and standard similarity functions. \cite{21} \\ \hline

Land surface & Noah Land-Surface Model & Developed jointly by NCAR and NCEP. \newline Includes root zone, evapotranspiration, soil drainage, and runoff, taking into account vegetation categories, monthly vegetation fraction, and soil texture. \cite{21} \\ \hline

Planetary boundary layer & Yonsei University scheme (YSU scheme) & Using the counter-gradient terms to represent fluxes due to non-local gradients. \cite{21} \\ \hline

Release location & Latitude: 21.64 \newline Longitude: 108.50 & At the Fangchenggang NPP. \\ \hline

\end{longtable}

\subsection{Result on temperature field} \label{sec:3.2}
\begin{figure*}
  \includegraphics[width=\linewidth, height=3.7in]{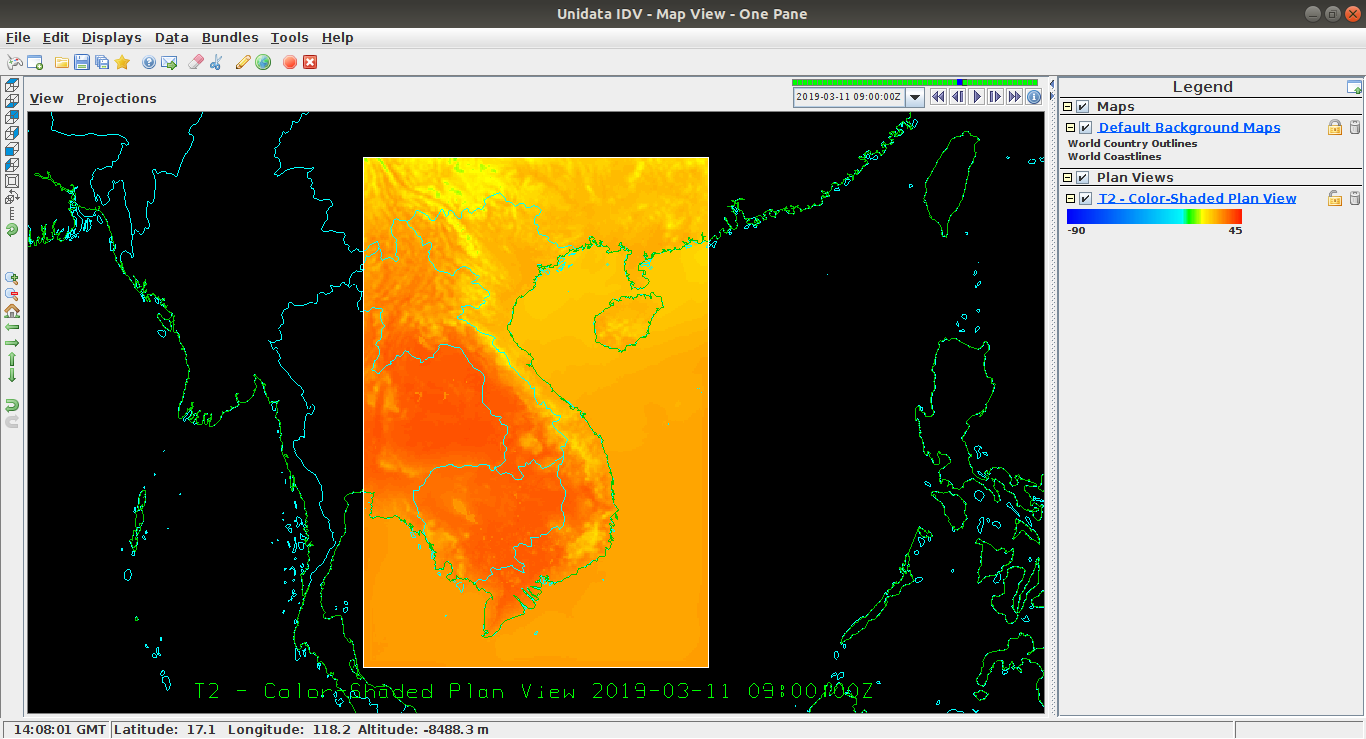}
  \caption{Visualization of the temperature field at 2m-height simulated at 11/03/2019 09:00:00 GMT. The unit is in $^{o}$C.}
  \label{fig:fig5}
\end{figure*}

\begin{figure*}
  \includegraphics[width=\linewidth, height=3.7in]{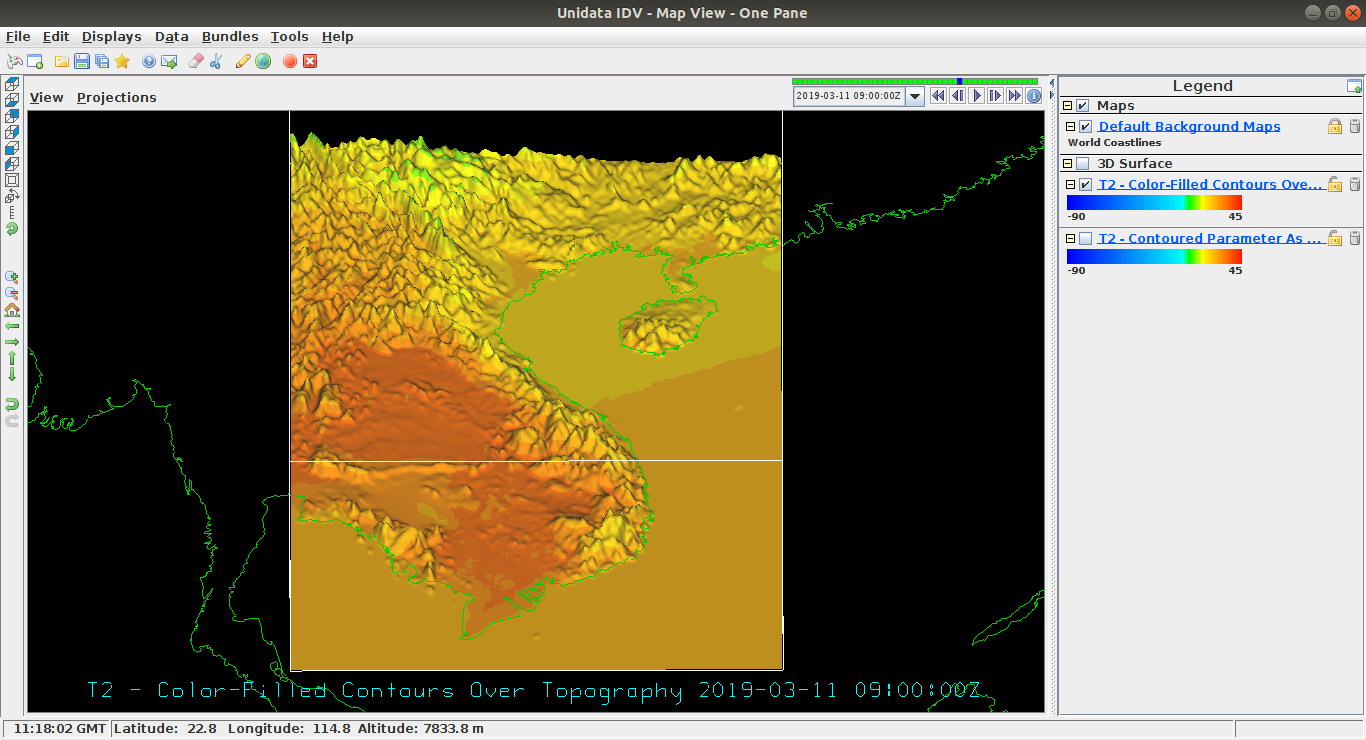}
  \caption{Hybrid visualization of the 2m-temperature field over typography simulated at 11/03/2019 09:00:00 GMT. This facilitates users a fusionable view on multiple variables.}
  \label{fig:fig6}
\end{figure*}

The temperature field is an important variable used for calculation of air density and turbulence parameters in PBL schemes, such as Monin–Obukhov length, turbulent kinetic energy (TKE), convective velocity, friction velocity \cite{17}. Although the dispersion process of aerosols is mainly occurred in the troposphere and stratosphere, in general, the greatest fraction of particle dispersion is in the tropopause level. Therefore, benchmarking of the meteorological simulation results is need to make comparisons with monitoring data at the tropopause level. However, due to we do not have monitoring data at the tropopause level and the temperature field could be vertical interpolated, so we decided to concern on the 2m-height temperature. \autoref{fig:fig5} shows the WRF simulation result on the temperature field at 2m-height in color-shaded mode and \autoref{fig:fig6} shows the temperature field at 2m-height over typography.  

\begin{figure}
    \centering
    \subfigure[]{\includegraphics[width=0.63\textwidth,height=2.2in]{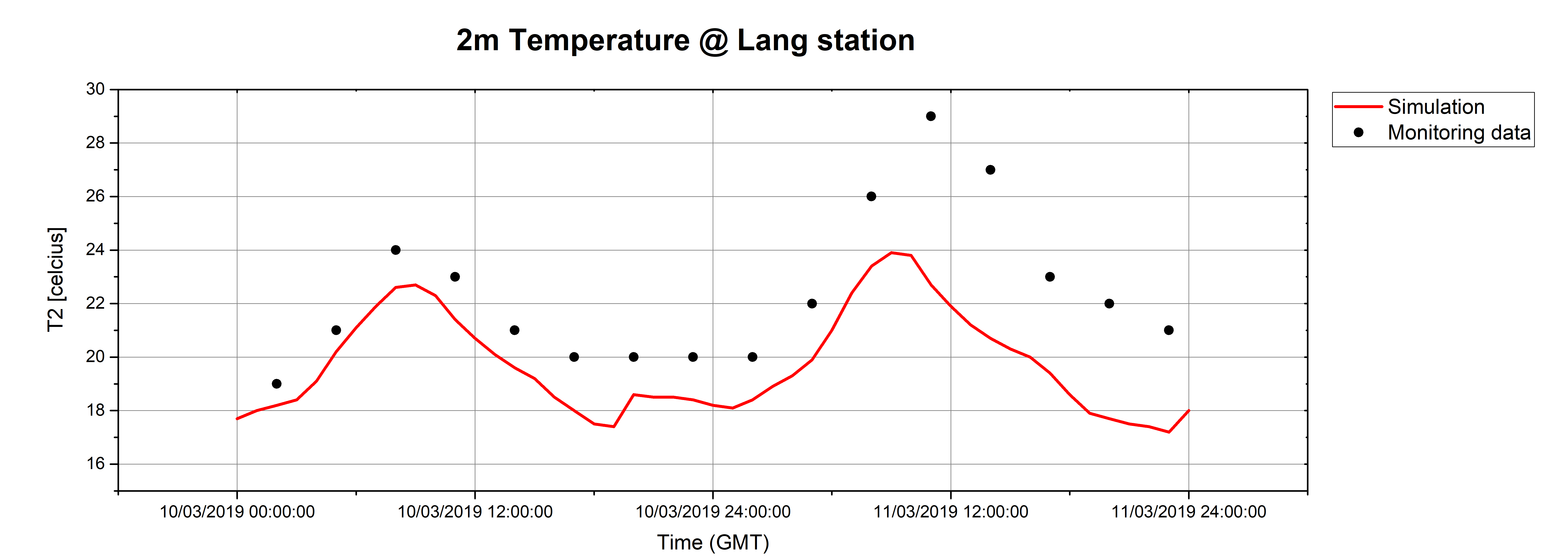}} 
    \subfigure[]{\includegraphics[width=0.34\textwidth,height=2.2in]{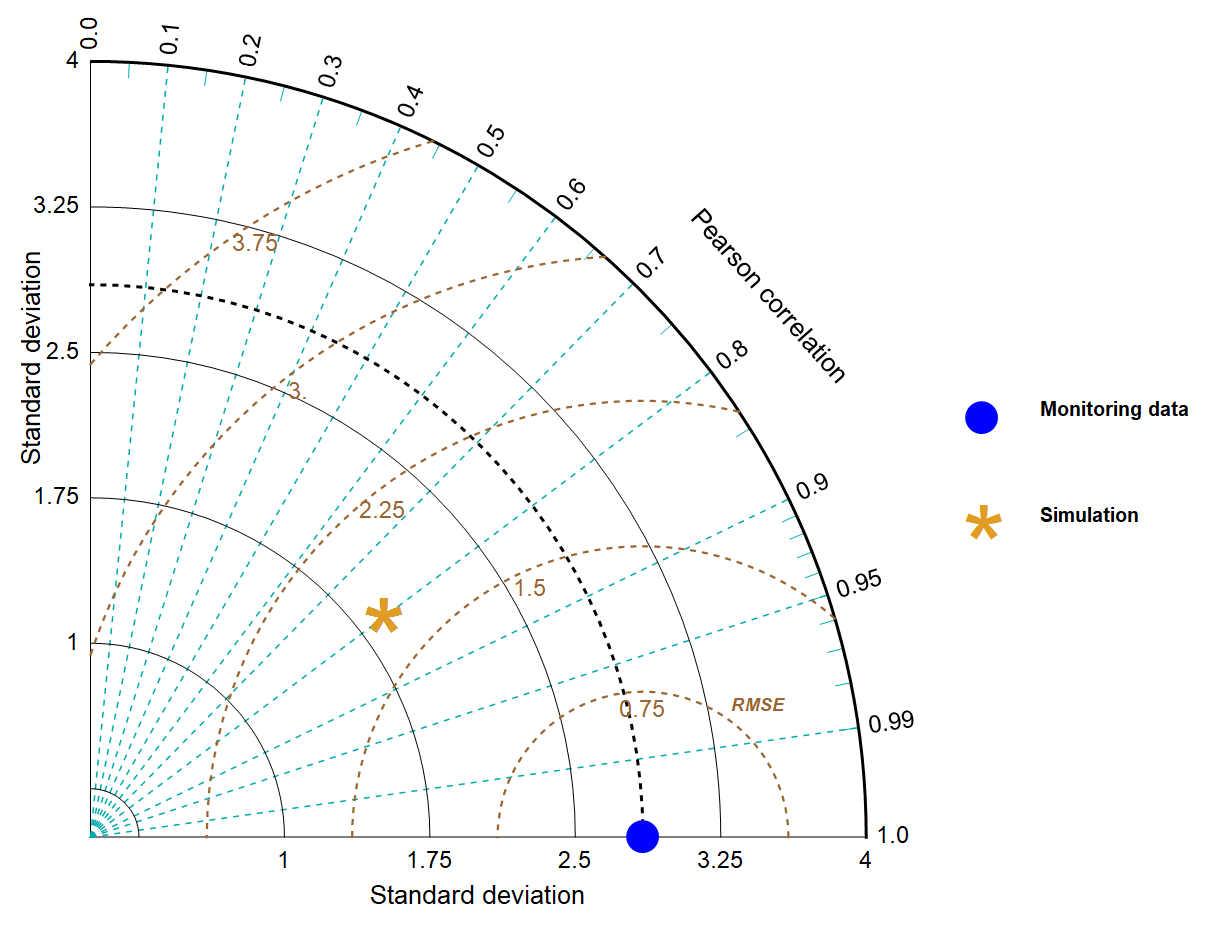}} 
    \caption{(a) Validation of simulation result on temperature field (red line) at the Lang station with observation data (black dot). (b) Taylor diagram to evalute simulation result and observed data on 2m- temperature.}
    \label{fig:fig7}
\end{figure}

The changing temperature in two days after 10/03/2019 00:00:00 GMT at the Lang station was simulated, the comparison with monitoring data has also been made and be illustrated in \autoref{fig:fig7}a.

The analysis shows that the temperature profile on \autoref{fig:fig7}a in both simulation result and observed data has a similar tendency, in general, the temperature of simulation is lower than the observed data. The highest point of monitoring data is 29$^{o}$C while with the simulation result is 24$^{o}$C. The difference between the simulation result and the observed data is more obvious when the time comes to the end of the second day. The relative merits of simulation result and observed data on 2m-temperature can be inferred from \autoref{fig:fig7}b, in which the correlation is 0.79, the standard deviation of observed data is 2.76 while with simulation result is 1.85, the Root Mean Square Error (RMSE) is 1.75.

\subsection{Result on wind field} \label{sec:3.3}

The WRF model uses the Arakawa C-grid, in which the horizontal and vertical wind components are defined at the side center of grid cells, while the FLEXPART-WRF model uses the horizontal coordinates is latitude/longitude and the vertical coordinate is the Cartesian terrain-following coordinate for saving computation time, the wind components are defined at the center of grid cells \cite{17}. There is a sub-routine of the FLEXPART-WRF model used to interpolate the wind field from WRF output to FLEXPART-WRF \cite{17}.

\begin{figure*}
  \includegraphics[width=\linewidth, height=3.7in]{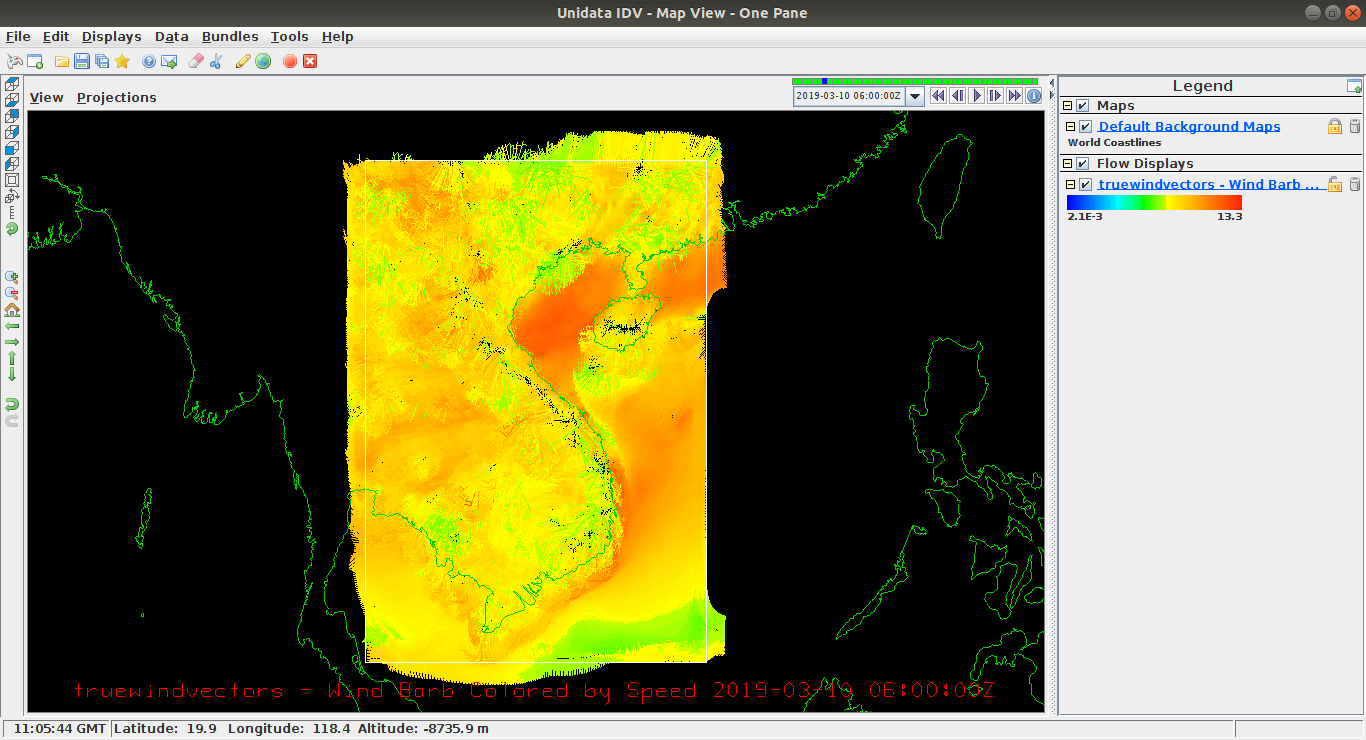}
  \caption{Representation of the wind bards with the speed (in m/s) is colored at the time 10/03/2019 06:00:00 GMT.}
  \label{fig:fig8}
\end{figure*}

\begin{figure}
    \centering
    \subfigure[]{\includegraphics[width=0.48\textwidth,height=2.5in]{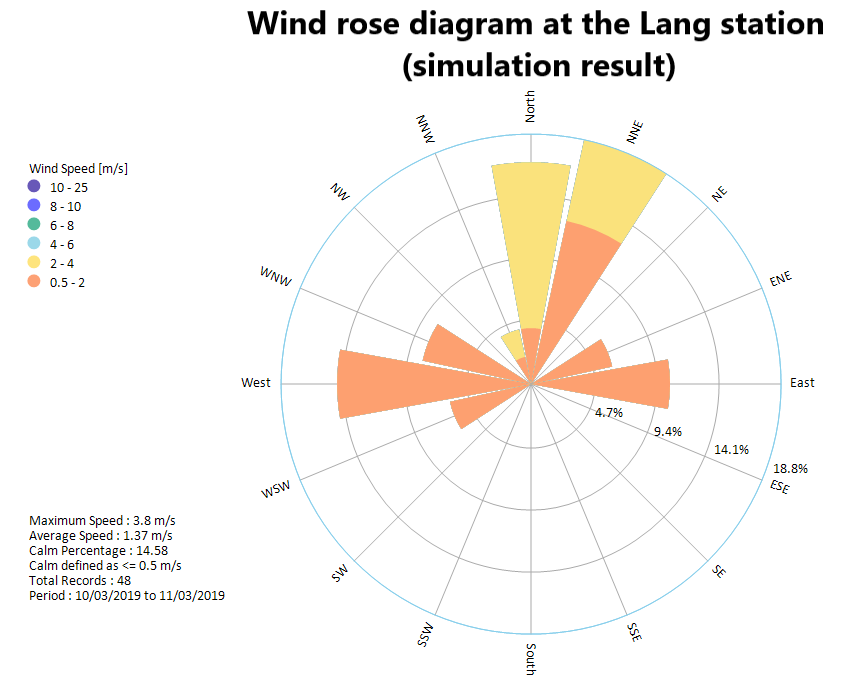}} 
    \subfigure[]{\includegraphics[width=0.48\textwidth,height=2.5in]{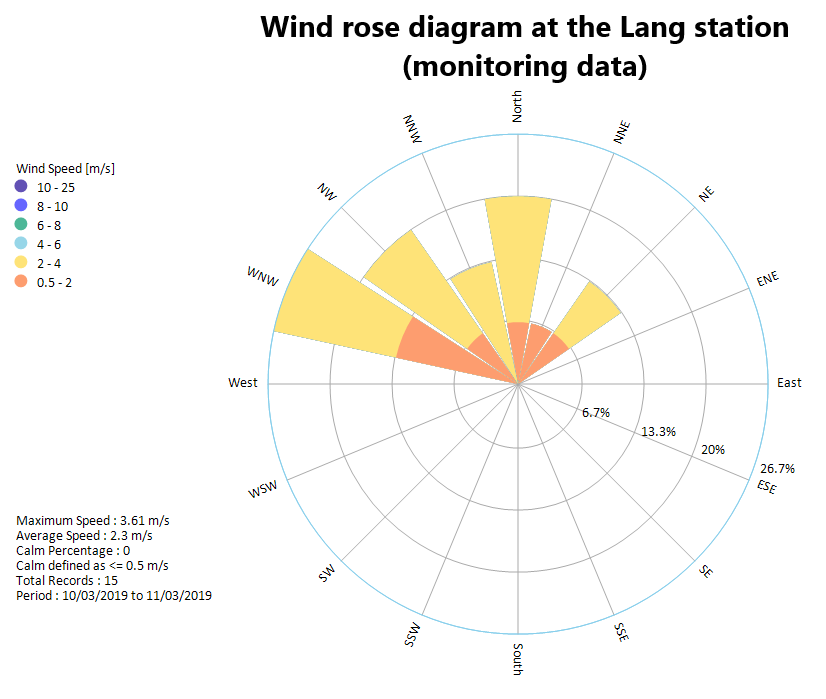}} 
    \caption{Wind rose validation at the Lang station with monitoring data in 48 hours after 10/03/2019 00:00:00 GMT. The simulation result shows the average wind speed was 1.37 m/s with the highest frequency wind direction was from NNE toward SSW. On the monitoring data, the average wind speed was 2.3 m/s with the highest frequency wind direction was from WNW toward ESE.}
    \label{fig:fig9}
\end{figure}

Wind field in WRF output could be interactively visualized in both its speed and direction (\autoref{fig:fig8}), here we presented the true wind vector of surface (10 m altitude) with the speed is colored. The wind rose analysis at the Lang station and validation with monitoring data are shown in \autoref{fig:fig9}.

\begin{figure}
    \centering
    \subfigure[]{\includegraphics[width=0.63\textwidth,height=2.2in]{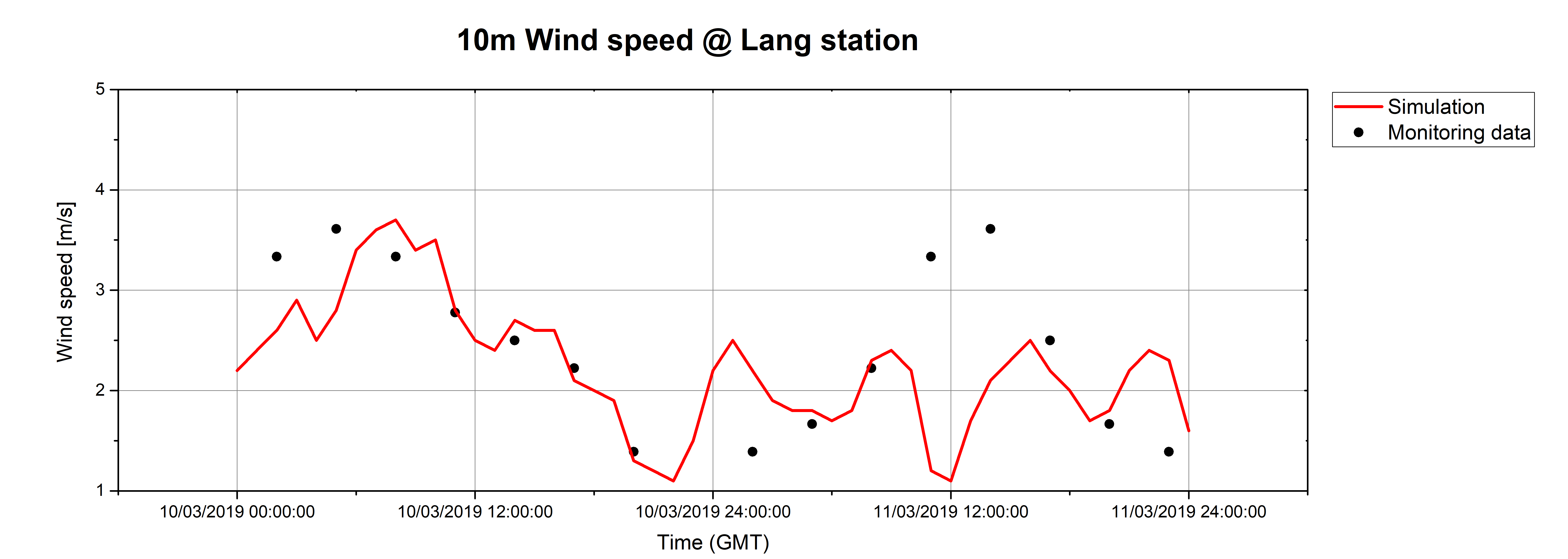}} 
    \subfigure[]{\includegraphics[width=0.34\textwidth,height=2.2in]{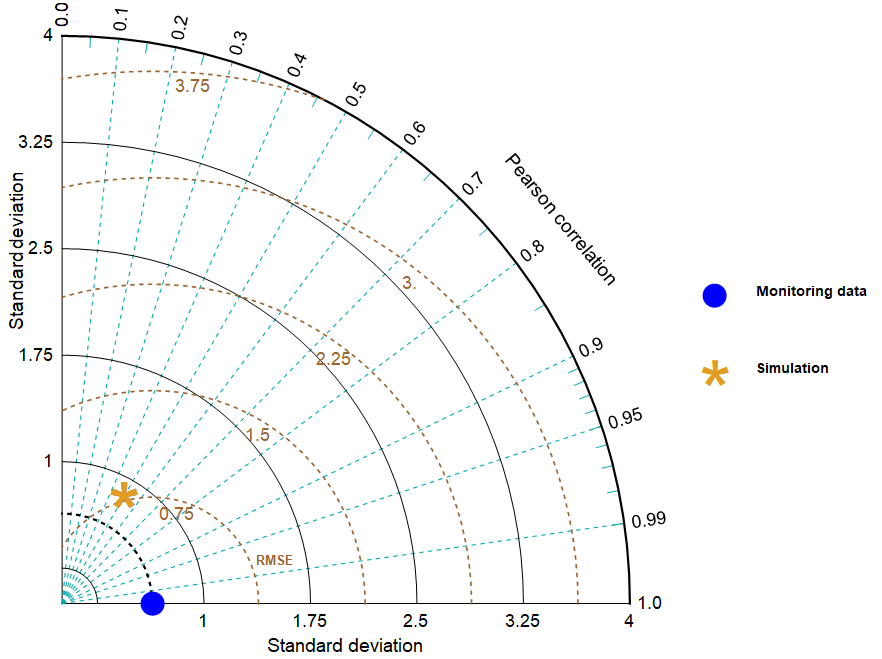}} 
    \caption{(a) Comparision of simulation result (red line) on wind speed at 10m-height with observation data (black dot) at the Lang station. (b) Taylor diagram for performance benchmarking of simulation result with observed data.}
    \label{fig:fig10}
\end{figure}

During this period, Hanoi was mainly influenced by the northeast monsoon, \autoref{fig:fig9} shows the highest frequency of wind direction in simulation result is from NNE toward SSW, but with monitoring data is from WNW toward ESE. There is no wind blowing from West, East, and South in monitoring data, however, simulation result shows there is 14.2\% of wind blow from the West and 9.7\% of wind blow from the East. A good correlation of the wind speed at 10m-height in simulation result and observed data in the first day (\autoref{fig:fig10}a), but in the second day there was a difference between these two data, especially around the time 11/03/2019 14:00:00 GMT. A Taylor diagram has taken into account for benchmarking simulation result with monitoring data (\autoref{fig:fig10}b), in which the RMSE is 0.8, the correlation is 0.49, the standard deviation of simulation result is 0.61 and with the observed data is 0.87.

\subsection{Result on surface pressure} \label{sec:3.4}
The surface pressure is used for calculation of turbulence parameters in PBL schemes \cite{17} and has a significant role in study the thermodynamical state of atmosphere, the 3D pressure field at an arbitrary level could be vertically interpolated. The WRF model uses the vertical coordinate system as terrain-following hydrostatic pressure based $\sigma$-levels (also called the ETA levels, which the lowest level corresponds to the surface level), while the FLEXPART-WRF uses the terrain-following Cartesian vertical coordinate system \cite{17}. FLEXPART-WRF uses a sub-routine to convert vertical wind components from WRF coordinate to FLEXPART-WRF coordinate by a correction factor of orography gradient \cite{17}. 

\begin{figure*}
  \includegraphics[width=\linewidth, height=3.7in]{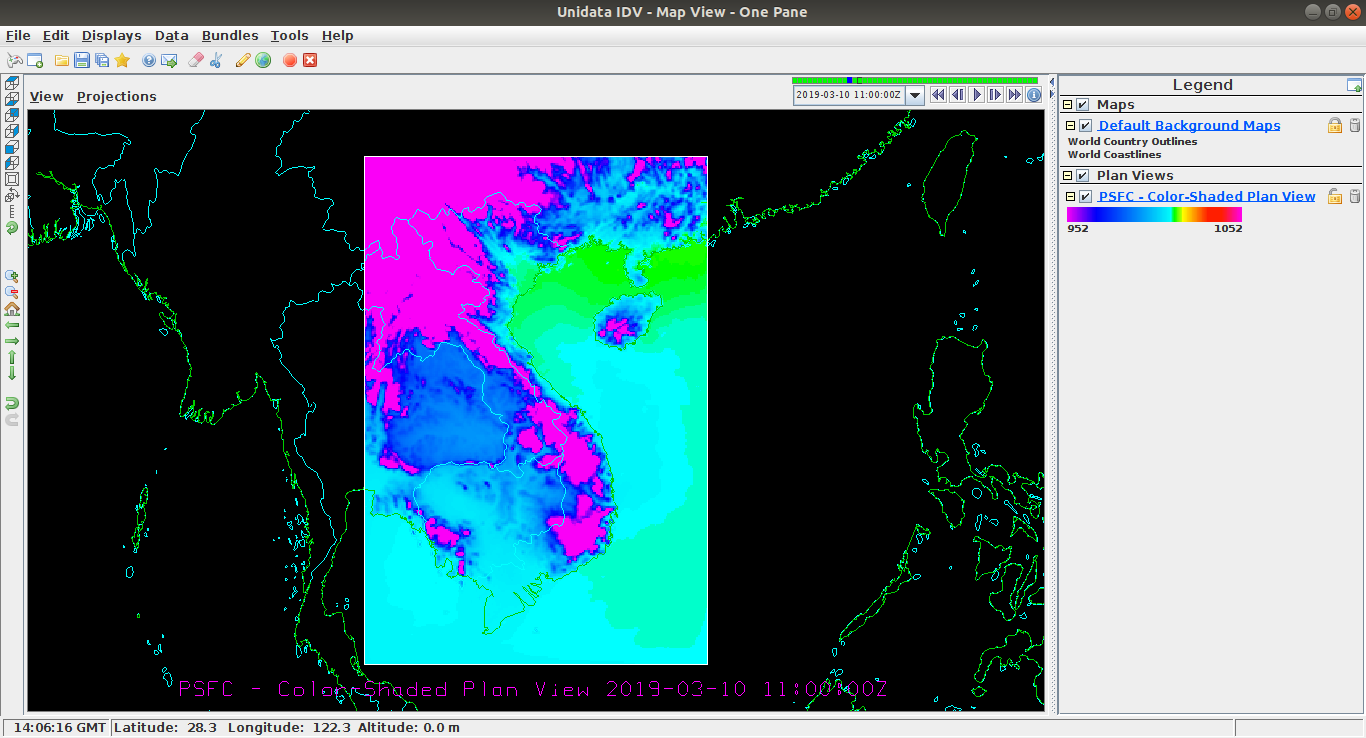}
  \caption{Visualization of the surface pressure on simulation domain at the time 10/03/2019 11:00:00 GMT. The unit is in hPa. }
  \label{fig:fig11}
\end{figure*}

\begin{figure}
    \centering
    \subfigure[]{\includegraphics[width=0.63\textwidth, height=2.2in]{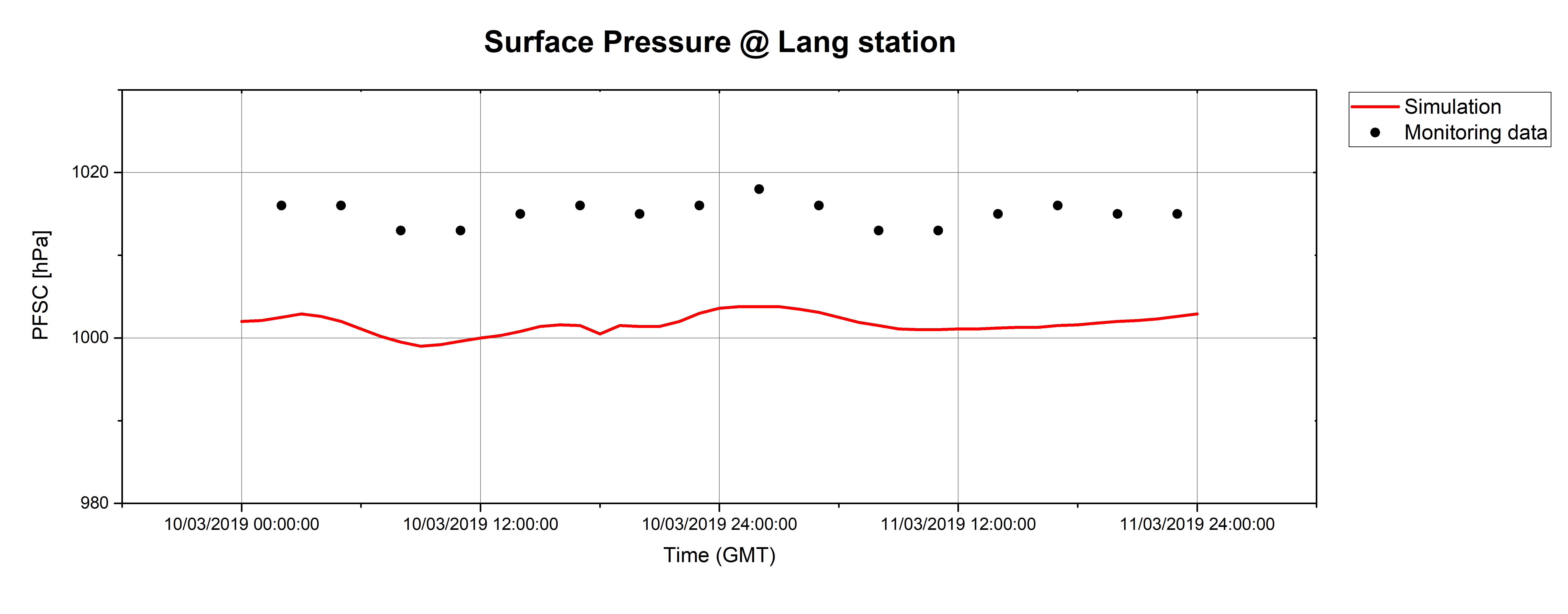}} 
    \subfigure[]{\includegraphics[width=0.34\textwidth, height=2.2in]{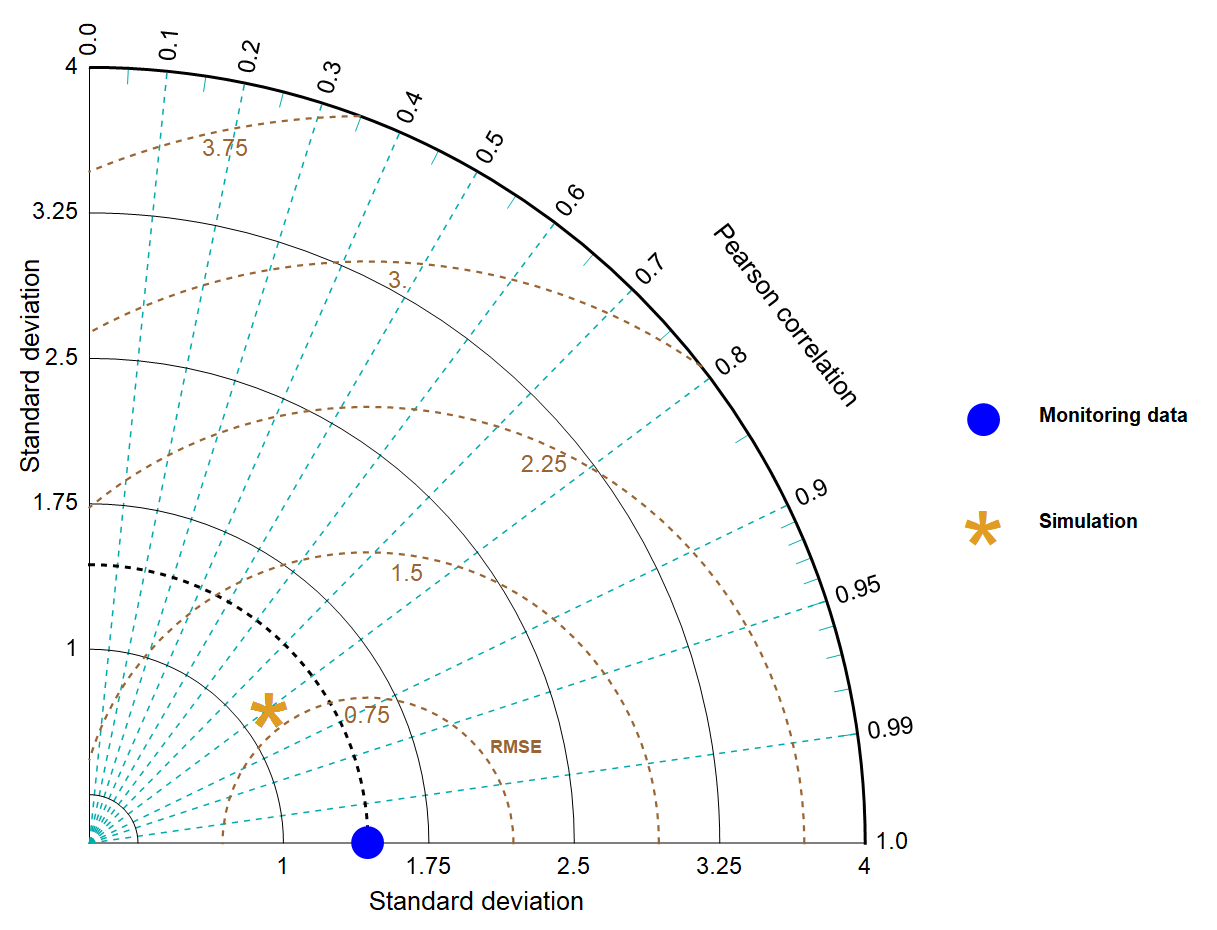}} 
    \caption{(a) Change of surface pressure in simulation result and observed data; (b) Taylor diagram for making comparision between simulation result and observed data.}
    \label{fig:fig12}
\end{figure}

The simulation result on surface pressure was visualized by IDV and was illustrated in \autoref{fig:fig11}. \autoref{fig:fig12}a shows the change of surface pressure in time at the Lang station in both simulation result and monitoring data. 

The surface pressure of simulation result and monitoring data has the same profile, in which the lowest point at 10/03/2019 09:00:00 GMT, and the highest point at 11/03/2019 01:00:00 GMT. However, the simulation gives result is lower than the observed data. A Taylor diagram has taken place to evaluate simulation result and observation on the surface pressure (\autoref{fig:fig12}b). The correlation is 0.81, the standard deviation of observed data is 1.39 while the simulation is 1.14, the RMSE is 0.86.

\subsection{Result on precipitation} \label{sec:3.5}

The FLEXPART-WRF uses two methods for calculation of the wet deposition are relative humidity-based scheme and precipitation-based scheme. The RH-based scheme is inherited from the FLEXPART for cloud definition (in-cloud or below-cloud) by setting the threshold of relative humidity is 80\% \citep{17} \citep{30}. The precipitation-based scheme is a new scheme developed by Seibert and Arnold, improved by Seibert and Philipp, and corrected by measurements from the Comprehensive Nuclear-Test-Ban Treaty Network \citep{17}. This scheme concern on the rainfall from microphysics process and cumulus process by extracting the precipitation rate from the WRF output data \citep{17} for wet deposition calculation, which expects to be better than the RH-based scheme.

\begin{figure*}
  \includegraphics[width=\linewidth, height=3.5in]{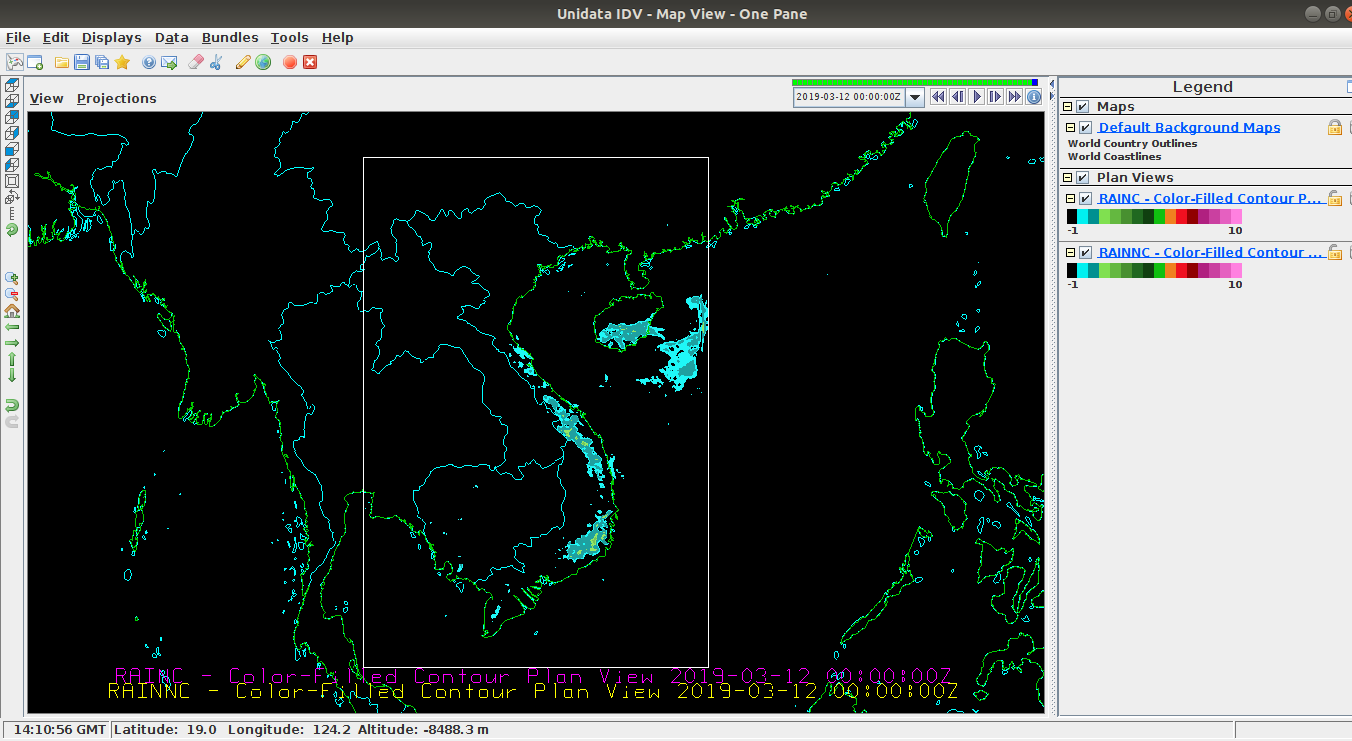}
  \caption{Visualization of the accumulated rainfall over simulation period. The result is sum of convective precipitation and non-convective precipitation.}
  \label{fig:fig13}
\end{figure*}

\autoref{fig:fig13} shows the accumulated precipitation over the simulation period (48 hours from 10/03/2019 00:00:00 GMT). Both simulation result and monitoring data at the Lang station shown that there is no rainfall during this time.

\subsection{Result on the dispersion of $^{137}$Cs} \label{sec:3.6}

\begin{longtable}{|p{0.45\linewidth}|p{0.45\linewidth}|}

\hline \hline 
\endfirsthead

\hline \hline
\endlastfoot

\textbf{a}\includegraphics[width=1\linewidth, height=10.0cm]{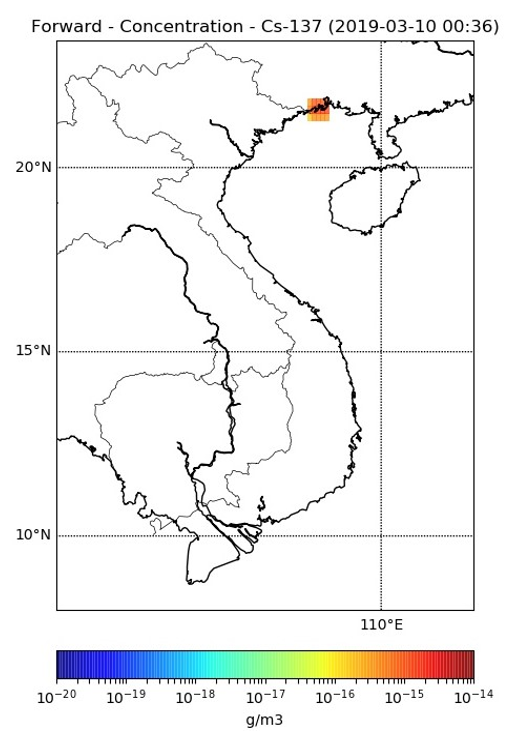} 
& \textbf{b} \includegraphics[width=1\linewidth, height=10.0cm]{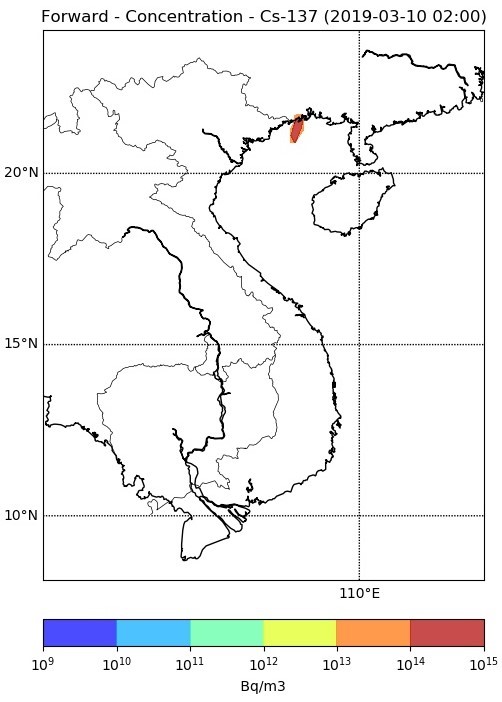}  \\ \hline 

\textbf{c} \includegraphics[width=1\linewidth, height=10cm]{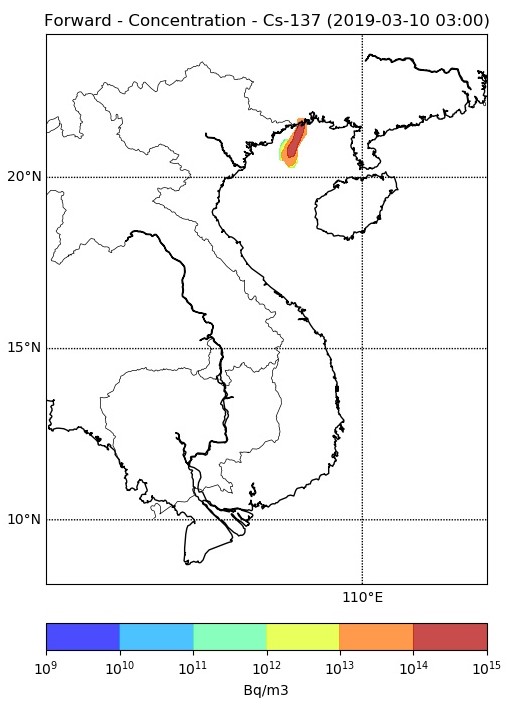} 
& \textbf{d} \includegraphics[width=1\linewidth, height=10cm]{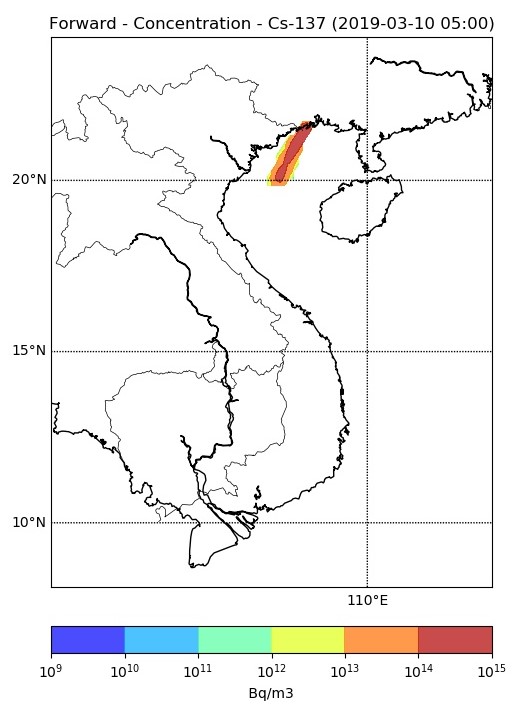}   \\ \hline 

\textbf{e} \includegraphics[width=1\linewidth, height=10cm]{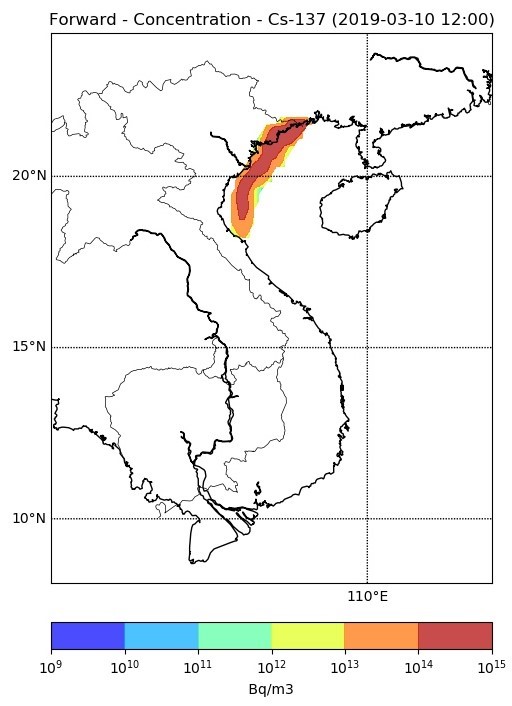} 
& \textbf{f} \includegraphics[width=1\linewidth, height=10cm]{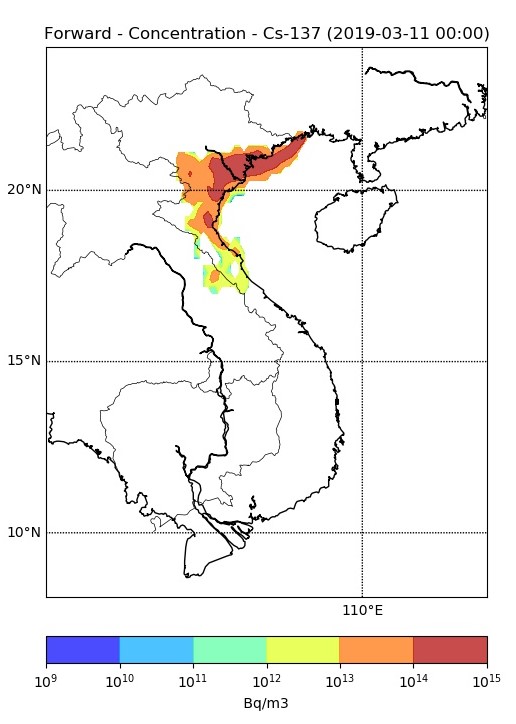}   \\ \hline 

\textbf{g} \includegraphics[width=1\linewidth, height=10cm]{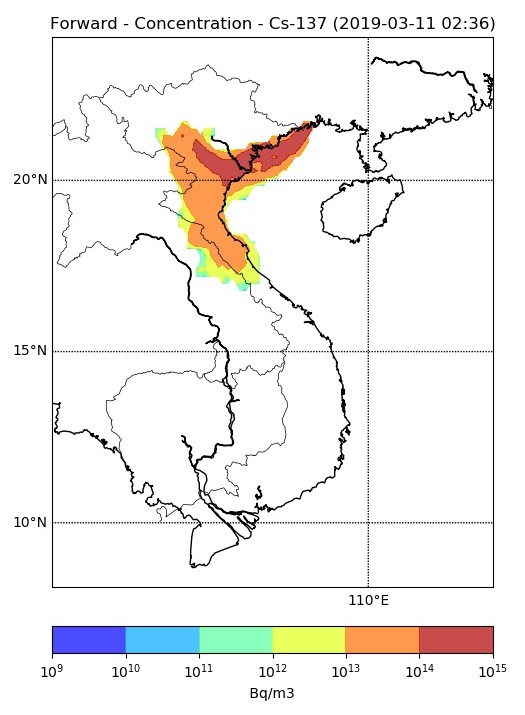} 
& \textbf{h} \includegraphics[width=1\linewidth, height=10cm]{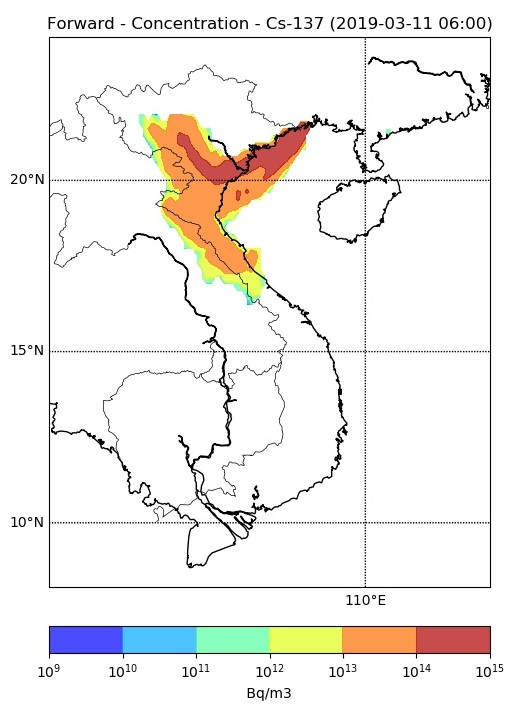}   \\ \hline 

\end{longtable}
\captionof{figure}{Trajectory of $^{137}$Cs dispersed from Fangchenggang NPP on the whole Vietnam territory by time evolution simulated by the GoldEnvSim.} \label{fig:fig14}

The output from FLEXPART-WRF was readout and processed by the Quicklook for visualization. \autoref{fig:fig14} describes the simulation result of $^{137}$Cs released from the Fangchenggang NPP and transported on the whole territory of Vietnam. After the first 5 hours (\autoref{fig:fig14}d), the influence of radioactive material on Vietnam was negligible. The dispersion direction toward the South-West by influence of the northeast monsoon.

24 hours after released (\autoref{fig:fig14}f), Hanoi capital was in the impact area of radioactive material, the concentration of $^{137}$Cs in Hanoi was at low level at this time. The wind blows from South China Sea caused radioactive material substances drifted to Laos and Northern Vietnam.

At 6 o’clock the next day (\autoref{fig:fig14}h), most of Northern Vietnam was in the influence area of radionuclide. There was a very high concentration in North Central of Vietnam, the concentration in Hanoi and other provinces nearby was above average level. On the contrary, from the 15th parallel of Vietnam downwards to South Vietnam, it was unaffected by radioactive material.

\section{Discussion and conclusions} \label{sec:4}

This work has made a number of significant contributions to the field of software development for computational environment applications, especially on the simulation of the dispersion of radionuclides and other forms of pollutants. For our first goal, we focused on solving an important problem, that is developed a GUI software to couple WRF module and FLEXPART-WRF module, which is helpful for end-users to run simulation more conveniently and consistently. A demonstration run is carried out to illustrate the software performance and show its applicable for emergency response and research purposes.

The WRF simulation results on temperature, wind speed, surface pressure are underestimated compared to the observed result taken from a first-class meteorological observatory. On the Taylor diagram, a simulation result that agrees well with the monitoring data has to lie nearest to the monitoring data on the x-axis, this means the simulation result has high correlation and low RMSE. The wind speed is the main factor that affect to the trajectory of the aerosol has the lowest correlation (0.49) and the lowest RMSE (0.8) compare to temperature and surface pressure. The temperature has a high correlation (0.79) and the highest RMSE (1.75). The highest accuracy of simulation results is on the surface pressure, which gives the highest correlation (0.81), the RMSE is 0.86, the simulation result and observed data have the standard deviations is near to each other. By evaluated parameters above, this simulation is concerned well suite to the observed data, however some different configurations on WRF schemes are needed to be tested in further research to get better accuracy.

For future development, backward simulation to predict the pollutant origin is planned to implement into the software. New advanced techniques for running WRF such as radar data assimilation (3D-Var), simulation with the moving-nest will be integrated into the GoldEnvSim.

\section{Discussion and conclusions} \label{sec:5}

This work is supported by the Military Institute of Chemical and Environmental Egineering Research Project in KC.AT Program, grant no 635/2017/HDKHCN. The authors would like to give special thanks to the NCAR for the WRF development and Prof. Andreas Stohl for the FLEXPART-WRF development.

\bibliography{mybibfile}

\end{document}